\documentclass[11]{article}
\usepackage[margin=1.2in]{geometry} 
\usepackage{amsmath,amsthm,amssymb,enumerate}
\usepackage{graphicx}
\usepackage{lscape}
\usepackage[english]{babel}
\usepackage[noend]{algpseudocode}
\usepackage{fixmath}
\usepackage{float}
\usepackage{tikz}
\usetikzlibrary{arrows}
\usepackage{verbatim}
\usetikzlibrary{calc}
\usepackage[shortlabels]{enumitem}
\usepackage{amsthm}
\usepackage{cite}
\usepackage{caption}
\usepackage{booktabs}
\usepackage{multirow}
\usepackage{lscape}
\usepackage{lipsum} 
\usepackage{authblk} 
\usepackage{titlesec}
\usepackage{forest}
\usepackage{placeins}
\usepackage{adjustbox}
\usepackage{multirow}
\usepackage{diagbox}
\usepackage{esvect}
\usepackage{algorithm}
\usepackage{algpseudocode}
\usepackage{algorithmicx}
\algdef{SE}[DOWHILE]{Do}{doWhile}{\algorithmicdo}[1]{\algorithmicwhile\ #1}%

\usepackage[utf8]{inputenc}
\usepackage[english]{babel}
\usepackage{csquotes}

\usepackage{ragged2e}

\usetikzlibrary{trees}

\usepackage[hyphens]{url}
\usepackage{hyperref}

\usepackage{qcircuit}

\newcommand{\bra}[1]{\langle{#1}|}
\newcommand{\ket}[1]{|{#1}\rangle}

\hypersetup{
	pdfauthor={Auth1, Auth2, Auth3},
	pdfkeywords={keywords, ...},
	pdfsubject={Security and Privacy; Cryptography},
	colorlinks = true,
	citecolor = blue,
	pdfstartview=Fit,
	breaklinks=true
}

\newcommand{\bi}{\begin{itemize}}
\newcommand{\ei}{\end{itemize}}

\algnewcommand\algorithmicformat{\textbf{Format:}}
\algnewcommand\Format{\item[\algorithmicformat]}

\title{\textbf{From Portfolio Optimization to Quantum Blockchain and Security: A Systematic Review of Quantum Computing in Finance}}

\date{}

\providecommand{\keywords}[1]{\textbf{Keywords }#1} 

\newcommand*{\affaddr}[1]{#1}
\newcommand*{\affmark}[1][*]{\textsuperscript{#1}}
\newcommand*{\email}[1]{\texttt{#1}}
\author{
	Abha Naik\affmark[1],
    Esra Yeniaras\affmark[2], 
    Gerhard Hellstern\affmark[3], 
    Grishma Prasad\affmark[4],
    Sanjay Kumar Lalta Prasad Vishwakarma\affmark[5] \\

	\affaddr{\affmark[1]VVM's Shree Damodar College of Commerce and Economics, Goa, India}\\
 \email{abhanaikimq@gmail.com}\\
	 \affaddr{\affmark[2]IT University of Copenhagen, Copenhagen, Denmark}\\
  \email{esye@itu.dk}\\
  \affaddr{\affmark[3]Center of Finance, DHBW Stuttgart, Herdweg  29, D-70174 Stuttgart, Germany}\\
  \email{gerhard.hellstern@dhbw-stuttgart.de}\\
\affaddr{\affmark[4]IIT, Bombay, India}\\
\email{grishmaprs@gmail.com}\\
\affaddr{\affmark[5] IBM Quantum, Almaden Lab, California USA}\\
\email{sanjay.vishwakarma@ibm.com}
}

\begin{document} 
\maketitle
\begin{abstract}
\justify
In this paper, we provide an overview of the recent work in the quantum finance realm from various perspectives. The applications in consideration are Portfolio Optimization, Fraud Detection, and Monte Carlo methods for derivative pricing and risk calculation. Furthermore,  we give a comprehensive overview of the applications of quantum computing in the field of blockchain technology which is a main concept in fintech. In that sense, we first introduce the general overview of blockchain with its main  cryptographic primitives such as digital signature algorithms, hash functions, and random number generators as well as the security vulnerabilities of blockchain technologies after the merge of quantum computers considering  Shor's quantum factoring and Grover's  quantum search algorithms. We then discuss the privacy preserving  quantum-resistant blockchain systems via threshold signatures, ring signatures, and zero-knowledge proof systems i.e. ZK-SNARKs in quantum resistant blockchains. After emphasizing the difference between the quantum-resistant blockchain and quantum-safe blockchain we mention the security countermeasures to take against the possible quantumized attacks aiming these systems. We finalize our discussion with quantum blockchain,  efficient quantum mining and necessary infrastructures for constructing such systems based on quantum computing. This review has the intention to be a bridge to fill the gap between quantum computing and one of its most prominent application realms: Finance. We provide the state-of-the-art results in the intersection of finance and quantum technology for both industrial practitioners and academicians.\\
\justify
\noindent\keywords{Portfolio optimization, Fraud Detection, Derivative Pricing, Risk Calculation, Monte Carlo, Quantum blockchain, Quantum-resistant blockchain, Digital signature algorithms, Post-quantum cryptography, Security, Privacy preserving blockchain, Quantum Computing}
\end{abstract}

\section{Introduction and related work}
\justify
In recent years, the topic of quantum computing has gained a lot of traction -  not only from physics or computer science domains but also from potential application areas. One of these areas where it is widely believed that quantum benefits could arise is finance. This is because the financial world is full of hard and computing-intensive problems and many of which have substantial potential to effectively exploit the opportunities provided by quantum computing. 

In finance, however, there are several topics, where the possibilities of quantum computing are being explored. Therefore, it is quite difficult to get a coherent overview of the possibilities. One of the first reviews addressing the connection between quantum computing and finance is \cite{Egger_2020}. The authors cover the topics of Monte Carlo Simulation, Optimization and Machine Learning. However, all contributing authors are employees of IBM and they concentrate on describing approaches where quantum hardware and software of IBM is used.

The review \cite{Bouland_2020} covers the topics Monte Carlo simulation, portfolio optimization, and machine learning and is aimed toward an audience of  financial professionals with no particular background in quantum computation. Again in \cite{ORUS2019} the same topics are schematically covered but more from a physicist's point of view.

On the other hand \cite{herman2022survey} provides a survey of the foundational quantum algorithms, and discusses the works in quantum finance in the fields of Stochastic Modelling, Optimization, and Machine Learning. The paper aims at providing a broad view  of concepts and discusses a wide range of quantum-finance related work.
 
In this work, our objective is to provide an update on the work going on in the quantum finance realm. It builds specifically upon the recent work of \cite{herman2022survey} and can be viewed as a summary of the research in the quantum community from 2021 to 2023. However, this paper contributes by discussing the works from an application point of view. We aim at providing a detailed account of works in the three fields with the most interest from the research and finance community. After a brief introduction to the basics of quantum computing and the most important algorithms, we consider the applications of the following field:  Portfolio Optimization (section (\ref{PortfOpt_Section})), Fraud Detection (section (\ref{Fraud_sec})) and Monte Carlo methods for derivative pricing and risk calculation (section (\ref{MonteCarlo_sec})). We also highlight the research of  quantum computing's application in the field of the blockchain (section (\ref{Blockchain_sec})).  In that sense \cite{NISTStatus}
\cite{cryptoeprint:2022/026}
\cite{FernndezCarams2020TowardsPB}
\cite{Torres2020PostQuantumLR}
\cite{Dragan} are the main works that are surveyed together with various other individual quantum resistant blockchain systems proposals including the quantum resistant privacy enhancing ones. We examined the quantum blockchain, quantum mining, together with discussing both obstacles and advantages of its real world constructions in finance.  Especially the blockchain topic which is regularly missing in most of the the discussions and reviews of quantum computing in finance is necessary to be analyzed well to get the full picture.

\section{Basics of quantum computing}
\justify
Quantum computing is a nascent and developing field. A lot of research and funding by government, public and private companies have been put into this emerging domain as it has the potential to solve NP-hard problems with great accuracy \cite{preskill2012quantum}. The current state of the art of quantum computers is Noisy, and the qubits are very low in number, resulting in inaccuracy \cite{9897404}. Scientists and researchers are working on designing fault-tolerant quantum computers and are developing error mitigation techniques that enhance accuracy and are trying to show a quantum advantage \cite{Preskill_2018}.

\section{Quantum Mechanics}
Quantum mechanics is a sub-field of physics that investigates how matter and energy behave at the atomic and subatomic scales. Unlike classical mechanics, which is the study of the motion of macroscopic objects, it is a basic theory that offers a description of the physical universe.

Early in the 20th century, the fundamental ideas of quantum mechanics were initially established to describe how atoms and the particles that make them up behave. It is a probabilistic theory, thus rather than forecasting the precise locations and velocities of the particles, it deals with the likelihood of finding them in certain states. Two of the most important features of quantum mechanics are the concepts of superposition and entanglement \cite{barletta2023introduction}.

\subsection{Superposition}
Superposition allows particles to exist in multiple states at the same time. In other words, the wave function of a quantum system can be expressed as a linear combination of wave functions of its individual states. 

Let there be two states $\ket{0}$, $\ket{1}$, the possible superposition of these states is 
   \begin{center} $\ket{\psi}$ = $\alpha\ket{0}$ + $\beta\ket{1}$ \end{center}
   \begin{center}
       $\ket{0}$ = $\begin{bmatrix}
        1 \\
        0  
    \end{bmatrix}$
   \end{center}

      \begin{center}
       $\ket{1}$ = $\begin{bmatrix}
        0 \\
        1  
    \end{bmatrix}$
   \end{center}
where $\alpha$ and $\beta$ are complex coefficients. The probability of measuring the system in the state $\ket{0}$ is then $\alpha$, and the probability of measuring it in the state $\ket{1}$ is $\beta$. The coefficients $\alpha$  and $\beta$ can be determined by applying the normalization condition. The superposition principle is a fundamental concept in quantum mechanics and has many important applications, such as in quantum computing and quantum cryptography.

\subsection{Entanglement}
The link between two or more quantum systems is known as entanglement, which is a key idea in quantum physics. When two or more quantum systems are intertwined, it is impossible to characterize one system's state in isolation from the others, and even if the systems are far apart, a measurement of one can quickly change the state of the others. Quantum entanglement is key in quantum communication and quantum cryptography.

\section{Quantum Computing}
Quantum computing is a new paradigm of computing that uses quantum bits or qubits which can simultaneously represent numerous states in contrast to classical computers. The underlying idea behind traditional computation and information is the bit. The quantum bit, also known as the Qubit, is the foundation upon which quantum computation and quantum information are based. The construction of the classical computer uses wires and logic gates while the construction of the quantum computer uses wires and elementary quantum gates. A traditional bit can only be a 0 or a 1. The fact that a Qubit can exist in a state other than $\ket{0}$ or $\ket{1}$ distinguishes it from conventional bits \cite{prashant2007study}.

A qubit's special characteristic enables quantum computers to carry out some types of calculations far more quickly than conventional computers. For instance, quantum computers excel at factoring huge numbers - a significant issue in cryptography. They can also be used to simulate intricate physical systems, solve optimization issues, and perform other jobs that call for the processing of a lot of data. It is to be noted that these qubits vary depending on the hardware used. For example, IBM uses superconducting qubits, Xanadu's quantum computers are built using photons and Microsoft with ion traps. 

The manipulation of qubits is done using quantum gates, which are analogous to classical logic gates. A bunch of quantum gates acting on a couple of qubits makes a quantum circuit. Quantum gates are basically divided into two types: Single qubit quantum gates, Multi qubit quantum gates.

\subsection{Single Qubit Quantum Gates}
Single-qubit quantum gates are quantum gates that operate on a single qubit. These gates can be combined to create more complex operations, and they are the building blocks of many quantum algorithms. Some of the most used single qubit quantum gates are as follows:
\subsubsection{X Gate}
The single-qubit quantum gate X, usually referred to as the "NOT" gate, is frequently employed in quantum computing. It is equivalent to a radian-sized rotation of the qubit state vector about the Bloch sphere's x-axis. The Bloch sphere is a three-dimensional sphere, with the north and south poles representing the quantum state
$\ket{0}$ and $\ket{1}$, respectively. The equator of the sphere represents the superposition of these two states, which can be represented as a linear combination of $\ket{10}$ and $\ket{1}$.

The NOT gate flips a bit's value from 0 to 1 or from 1 to 0 in traditional computing. Similar to this, applying an X gate to a qubit in the states $\ket{0}$ or $\ket{1}$ will change it into the corresponding states. For this reason, the X gate is frequently referred to as the NOT gate's quantum analog.\\
The matrix representation of the X gate is as follows:
\begin{center}
    $\begin{bmatrix}
        0 & 1 \\
        1 & 0 
    \end{bmatrix}$
\end{center}

\subsubsection{H Gate}
The H gate, also known as the Hadamard gate, is a single-qubit quantum gate that is commonly used to create the superposition of two qubits. It corresponds to a rotation of the qubit state vector around the x-axis of the Bloch sphere by $90$ degrees, followed by a rotation around the z-axis by $180$ degrees. When the H gate is applied to a qubit in the state $\ket{0}$, it transforms it into a superposition of $\ket{0}$ and $\ket{1}$ states, represented by the state $1/\sqrt{2}[\ket{0} + \ket{1}]$. Similarly, when the H gate is applied to a qubit in the state |1⟩, it transforms it into the superposition $1/\sqrt{2}[\ket{0} - \ket{1}]$.
The matrix representation of the X gate is as follows:\\
\begin{center}
    $1/\sqrt{2}
    \begin{bmatrix}
        1 & 1 \\
        1 & -1 
    \end{bmatrix}$
\end{center}

\subsection{Multi Qubit Quantum Gates}
Multi-qubit quantum gates are quantum gates that operate on two or more qubits simultaneously. These gates are necessary for performing more complex quantum operations and implementing quantum algorithms. The two most commonly used multi-qubit quantum gates are the Controlled-NOT (CNOT) and SWAP gates. It is to be noted that the Hadamard gate and the CNOT gate form an entangled pair of qubits. All quantum algorithms in the circuit-based approach use the same basic two-qubit gates such as the Toffoli gate, CNOT gate, and SWAP gate.

\section{Quantum Algorithms}
Quantum algorithms are algorithms designed to run on a quantum computer. These algorithms are based on the principles of quantum mechanics. They are designed to take advantage of the unique properties of quantum systems, such as superposition and entanglement, to solve problems faster than classical algorithms \cite{Zhang_2022}. Shor's algorithm, which is used to factor enormous numbers and has important implications for encryption, is also one of the most well-known quantum algorithms. The RSA encryption, which is currently employed for securing internet communication, may be broken using Shor's algorithm. Grover's method, which is used to scan unsorted databases and has the potential to speed up various optimization issues, is another powerful quantum algorithm. In machine learning and artificial intelligence, quantum algorithms may be utilized to analyze huge datasets and carry out intricate computations. Beyond these particular uses, quantum algorithms are anticipated to have a considerable influence on industries including pharmaceutical research, materials science, and financial modeling. For instance, using quantum algorithms to model molecular behavior might hasten the search for novel materials and medicines \cite{santagati2023drug}. Quantum algorithms can be utilized in finance for risk assessment and portfolio optimization \cite{herman2022survey}. In this section, we will discuss some of the important quantum algorithms that are used in quantum finance.

\subsection{Grover's Algorithm}
Grover's algorithm is a quantum algorithm that is used to discover a specific item in a list of unstructured data or search an unsorted database. It was developed in 1996 by Lov Grover and is regarded as one of the most significant quantum algorithms. Grover's technique is exponentially quicker for large data sets because its efficiency is substantially higher than that of traditional algorithms, which must scan the whole database one item at a time. Grover's approach uses the amplitude amplification phenomenon and offers a quadratic speedup, which indicates that it needs fewer operations to search a database of N unsorted items than traditional algorithms, which need O(N) operations. For more details regarding the steps of  Grover`s algorithm please refer to Section 9.2.2.

The algorithm starts with a superposition of all possible solutions to the problem, represented by quantum states. The superposition is then iteratively modified using a quantum oracle($U_{f}$) and a quantum diffusion operator($U_{\psi}$). Grover's oracle is a black box that encodes the marked item or the item one is interested to find in the unsorted data set by flipping the state while, Grover's diffusion operator is used in amplitude amplification of the marked state thereby increasing the probability of occurrence during measurement. By repeating this process multiple times, the algorithm narrows down the search to the solution with high probability \cite{grover1996fast}.\ 

Portfolio optimization in traditional finance is a computationally demanding operation that requires looking through a large number of potential portfolios. On a quantum computer, this search can be carried out significantly more quickly with the help of Grover's method. It is intended to represent the portfolio optimization issue as a search problem, with the ideal portfolio as the goal item. The best portfolio is then identified by the quantum oracle, which also flips its amplitude's phase. The method is then repeated numerous times, which reduces the search to the most likely ideal portfolio. Credit scoring is one other area in which Grover's method may be used in finance. The algorithm may be applied to credit data to look for trends and locate clients at high risk of default. This can reduce the risk of default and help lenders make better lending decisions.\

\subsection{Quantum Fourier Transform}

The Quantum Fourier Transform (QFT) is a quantum algorithm that performs a Fourier transform on a quantum state. The Fourier Transform (FT) is one of the most useful mathematical tools in modern science and engineering. The FT is especially useful when we have something with underlying periodicity. We then remove the high-frequency components (the noise) from the spectrum and inverse-FT to give a clean set of data. FT allows us to extract the underlying periodic behavior of a function. Many quantum algorithms, such as Shor's algorithm for factoring big numbers and quantum phase estimation, use the QFT as a crucial subroutine. Shor's approach is based on the QFT's ability to determine a function's periodicity.\\

Let there be a quantum state $\ket{\psi}$ such that,\\

 $$ \ket{\psi} =\sum_{j=0}^{N-1} a_{j} \ket{j} = \begin{pmatrix}
a_{0}\\
a_{1}\\
a_{2}\\
.\\
.\\
.\\
a_{N-1}

\end{pmatrix}$$ \\

Quantum Fourier Transform is applying Discrete Fourier Transform to a quantum state. Applying DFT to the above quantum state we get the following state, \\

$$ F \ket{\psi} =\sum_{k=0}^{N-1} b_{k} \ket{k}$$,\\
$$where, b_{k}=1/\sqrt{N} \sum_{j=0}^{N-1} a_{j} e^{2{\pi}ijk/n}$$\\

A series of quantum gates, including the Hadamard gate, phase shift gates, and controlled phase shift gates, can be used to implement the quantum field theory (QFT). All possible values of y are superimposed using the Hadamard gate, and the complex phase factors are applied to the superposition using the phase shift gates and controlled phase shift gates. The QFT is a powerful algorithm that can be used to speed up many computations in quantum computing, including quantum simulation, quantum chemistry, and optimization problems. It is also a fundamental building block for many other quantum algorithms, including Shor's algorithm for factoring large numbers \cite{kashani2022quantum}.

\subsection{Quantum Phase Estimation}
The eigenvalues of a unitary operator are estimated using the quantum procedure known as quantum phase estimation (QPE). The values for which the equation $A\ket{\psi}=\lambda\ket{\psi}$ holds, where A is an operator and $\ket{\psi}$ is an eigenvector, are known as eigenvalues. Estimating eigenvalues is a critical step in many quantum algorithms, including Shor's technique for factoring big numbers. The quantum state $\ket{\psi}$ used in the QPE method is an eigenvector of a unitary operator U with an unidentified eigenvalue. Following that, a sequence of Controlled-U gates is used, each of which is controlled by a qubit in a superposition of states. With each repetition, more qubits are utilized in the superposition, allowing the algorithm to estimate the eigenvalue more precisely.\\

The Quantum Fourier Transform (QFT) is a subroutine used by the recursive QPE method. The approach can extract the phase information about the eigenvalue by performing a Fourier transform on the superposition of states using the QFT \cite{svore2013faster}. Then, in order to estimate, the QPE method measures the qubits in the superposition. There are several possible uses for Quantum Phase Estimation (QPE) in quantum finance. Pricing financial derivatives, a computationally demanding operation that entails predicting the eigenvalues of large matrices, is one of the most promising applications. In traditional finance, Monte Carlo simulations or other numerical techniques are commonly used to price financial derivatives. On a quantum computer, this computation can be carried out considerably more quickly with the aid of QPE. Portfolio optimization is one more area in which QPE might be used in finance. The eigenvalues of the asset covariance matrix may be estimated using the technique, which can aid in choosing the best portfolio. Investors can reduce the risk of loss and make better-informed portfolio allocation decisions as a result of this.\\

\subsection{Shor's Algorithm}
Shor's algorithm is a quantum algorithm that is effective in efficiently factoring big composite numbers \cite{Shor_1997}. This method has the name of the 1994 inventor, mathematician Peter Shor. The algorithm is crucial because it has substantial effects on computer security and cryptography. The difficulty of factoring huge numbers is the foundation for the security of many contemporary encryption methods. Large-number factoring is a very time-consuming job in traditional computing that can take years to accomplish. Modern encryption schemes are seriously threatened by the speed with which Shor's algorithm may resolve this issue on a quantum computer.\\
Shor's algorithm is composed of two parts. The first part of the algorithm turns the factoring problem into the problem of finding the period of a function and may be implemented classically. The second part finds the period using the quantum Fourier transform and is responsible for the quantum speedup. The key step in Shor's algorithm is quantum phase estimation, which is used to estimate the period of a function. This step is performed using a quantum register that is used to estimate the phase angle of an eigenvalue of a unitary operator. The algorithm then uses this estimate to compute the factors of the composite number. Shor's algorithm has important implications for computer security and cryptography. It reveals how many of the encryption techniques currently in use to protect sensitive data may be cracked using quantum computing. Researchers are creating new encryption technologies that can withstand quantum assaults in response to this threat. For more details regarding the steps of  Shor`s algorithm please refer to Section 9.2.2.\\

\subsection{Variational Quantum Algorithms}
Variational quantum algorithms (VQAs) are a type of quantum algorithm that searches for a solution to a problem using methods from classical optimization. VQAs are made to run on near-term quantum computers, which have a low number of qubits and a high error rate. The basic goal of VQAs is to minimize the use of quantum resources by using quantum hardware as a co-processor for a conventional optimization procedure. The fundamental idea behind VQAs is to begin with a quantum circuit that is parameterized by a few different variables, or the "ansatz". The ansatz circuit is a quantum circuit with certain configurable parameters and a fixed structure. These variables are selected to maximize the circuit's output and minimize a certain cost function, usually a classical cost function \cite{Cerezo_2021}.\\

The process of running the ansatz circuit on a quantum computer, measuring the result, and then updating the parameter values - is how the optimization is carried out. Until the cost function is minimized or the desired precision is attained, this procedure is repeated \cite{Cerezo_2021}. Numerous issues, including optimization issues, machine learning challenges, and quantum chemistry simulations, have been successfully solved with VQAs. For many of these issues, especially those that are challenging to answer with classical computers, VQAs have the potential to offer large speedups over traditional techniques.\\

Quantum Approximate Optimisation Algorithm (QAOA) is a quantum optimization technique with high potential in the area of quantum finance \cite{canabarro2022quantum}. With the use of a series of parameterized quantum gates, the QAOA variational algorithm attempts to solve an optimization issue as closely as possible. Portfolio optimization is one of the most intriguing uses of quantum computing in finance. The practice of dividing the capital investments among the members of a group of assets in order to maximize returns while lowering risk is known as portfolio optimization.\\

The portfolio optimization problem may be solved using QAOA by determining the ideal set of weights for a given collection of assets. A quantum circuit that encodes the weights of the assets is subjected to the QAOA algorithm, and the resulting cost function is measured. The best weights for the given set of assets are then determined by optimizing the cost function using conventional optimization techniques \cite{Zhou_2020}. Other financial optimization issues, such as credit risk evaluation and option pricing, can be resolved using QAOA in addition to portfolio optimization. These applications, however, are still in the early phases of development and need more study.\\

In conclusion, QAOA has the potential to be a very effective tool in quantum finance for resolving challenging optimization issues that are beyond the capabilities of classical computers. The application of QAOA and other quantum optimization algorithms in finance is anticipated to grow as quantum technology continues to advance.

\section{Portfolio Optimization}
\label{PortfOpt_Section}

A portfolio is a collection of financial assets from the same or different asset classes built to achieve a certain purpose. Financial assets include assets like gold, stocks, bonds, etc. The goal of the portfolio can be to earn aggressive or moderate returns, minimize risk or have a balanced approach toward risk and returns. An efficient portfolio is one that achieves maximum returns for a given level of risk. Portfolio Optimization is a critical element of the Portfolio Selection and Management process which allows a portfolio manager to make the best selection among the available portfolios amidst dynamic market conditions like fluctuating market prices, changing interest rates, and dynamic legal and political scenarios.

In \cite{Markowitz1952}, the author proposed an investment model which greatly influenced and encouraged further investigations in the field of portfolio optimization. The Markowitz Portfolio Model provides us with the Capital Market Line (CML) which represents the risk-return trade-off in the capital market and is formulated as below:
\begin{equation*}
   r_{p}= i_{rf} + \frac{(r_{m}- i_{rf})\sigma_{p}}{\sigma_{m}}
\end{equation*}
  where $r_{p}$ = Expected Portfolio Return,
  
  $i_{rf}$ = Risk-free Rate of Interest,
  
  $r_{m}$ = Market Portfolio Return,
  
  $\sigma_{p}$ =  Standard Deviation of Portfolio
  
  $\sigma_{m}$ = Market’s Standard Deviation

This model aims at reducing the risk by diversifying the portfolio and accounting for risk-return trade-off thus providing the investors with an efficient portfolio based on their risk appetite.

\subsection{Previous Survey Works}
Each asset class in a portfolio optimization problem is assigned a weight. The assets are chosen based on factors like risk, return, average maturity, liquidity, etc.\cite{herman2022survey} discusses the portfolio optimization problems in two categories based on their formulations: Convex and Combinatorial Formulations. 
Portfolio optimization techniques \cite{Gunjan2022} have evolved from techniques like mean-variance \cite{Durand1960}, variance with skewness \cite{SAMUELSON1975}, Value-at-Risk Jorion \cite{Jorion1996ValueAR, Wipplinger2007}, Conditional Value-at-Risk\cite{Rockafellar2000OptimizationOC}, Mean-absolute deviation \cite{Konno1991MeanabsoluteDP} and Minimax \cite{Park1998AMP} to more advanced heuristic and meta-heuristic based methods. Now evolutionary algorithms and Swarm Intelligence have become popular choices for portfolio optimization. Apart from the above-listed classical approaches, the industry is also exploring several quantum and quantum-inspired algorithms for portfolio optimization. Being an optimization problem, Portfolio Optimization also suffers from the “curse of dimensionality” \cite{Kuo2005LiftingTC,Bellman1956}. Here the dimensionality of the data increases rapidly, which causes the volume of the data to increase as well. This results in the data becoming increasingly scattered and difficult to cluster. Quantum Computing holds the promise to tackle the problem due to its ability to handle larger computations faster than classical computing. 

Each asset class in a portfolio optimization problem is assigned a weight, and assets are chosen based on factors like risk, return, average maturity, liquidity etc.\cite{herman2022survey} discusses the portfolio optimization problems in two categories based on their formulations: Convex and Combinatorial Formulations. 

The combinatorial formulations are the ones that use integer optimization that use only binary optimization problems. An integer optimization problem is a mathematical optimization or feasibility program in which some or all of the variables are restricted to be integers and a binary optimization problem is the variant of an integer optimization that uses only 0s and 1s for variables. The financial optimization problems can be converted to a Quadratic unconstrained binary optimization (QUBO) problem\cite{enwiki:1146342738} which can then be linked to an Ising Hamiltonian. Finding the ground state of the Ising Hamiltonian is equivalent to finding the optimal solution to our QUBO problem i.e our optimal portfolio. Researchers have used QUBO formulations and included constraints of budget and correlational considerations to find whether to include an asset based on risk and also whether to have a long or short position. The binary variable-constrained mean-variance portfolio optimization problems also include a desired level of return. Researchers have also been able to show speedups and dynamic decision-making at multiple time steps. The quantum algorithms used for these optimization problems are suited for the NISQ era and generally rely on quantum annealing\cite{PhysRevE.58.5355} due to the larger number of qubits as compared to the gate-based models. The Convex Formulations use the quantum algorithms that employ convex optimization \cite{enwiki:1146817250}. The mean-variance portfolio optimization problem can be reformulated as a convex problem. The solution to these problems tells users the proportion of the amount to be invested in the asset instead of whether or not to include the asset in the portfolio. 

\subsection{Recent Developments}
\label{Rec_Dev}

The two major computation models being used by the researchers for portfolio optimization are Quantum Annealing (eg. D-Wave's systems) and Gate-based models (eg. IBM's devices). Quantum Annealing is suitable for certain problems like optimization problems, while Gate-based models are more universal with respect to the problems that can be solved using the model. However, Quantum annealing systems have been able to achieve more stable qubits than the Gate based systems, however, these qubits face the issue of low connectivity.

In \cite{Elsokkary2017FinancialPM}, a Markowitz Portfolio Optimization problem for stocks from the Abu Dhabi Securities Exchange was formulated as a QUBO and solved using D-Wave’s simulator for its  quantum optimizer-QBSOLV.The researchers made use of the Chimera architecture of the D-Wave system.

In \cite{Venturelli2019}, the researchers discuss a quantum-classical hybrid approach to the Markowitz portfolio optimization problem using the 2000Q D-Wave Quantum Annealer, wherein it was found that the best time-to-solution return, as a function of a number of variables, was obtained by seeding the quantum annealer with a solution candidate found by a greedy local search (classical component) and then performing a reverse quantum annealing protocol (quantum component). The Time-To-Solution is defined as the expected number of independent runs of the method in order
to find the ground state with a given probability (confidence level). The application in the discussion was- a fund of funds manager selecting a suitable fund from a universe of funds using a particular trading strategy. This was able to beat the pure classical solution, Genetic Algorithm, over the problem sizes of 42 - 60 assets with respect to time-to-solution. The optimized reverse annealing protocol was 100 times faster than the corresponding forward quantum annealing on average. The researchers also suggested that the transition to the Pegasus from the Chimera architecture will improve the performance by great magnitudes. An enhancement in the embedding process is very essential to improve the performance of D-Wave devices. An improvement in the QUBO formulations of \cite{Venturelli2019} was proposed in \cite{mattesi2023financial}, which allowed the investor to decide the optimal amount of investment in each asset instead of focusing only on the selection of optimal assets by discretizing the problem variables and continuous portfolio weights.

In \cite{cohen2020portfolio} and \cite{cohen2020portfolio2}, the authors performed the portfolio optimization of 40 and 60 US stocks respectively using D-Wave’s 2000 Quantum Annealer, following a buy-and-hold- strategy. The results were compared against classical algorithms. The researchers were positive that the results indicate that quantum advantage is possible as the number of assets under consideration increases. They developed CQR(Chicago Quantum Ratio) and CQNS (Chicago Quantum Net Score) as an improvement over the Sharpe Ratio as the D-Wave quantum annealer can only handle linear quadratic equations and not ratios.

\begin{equation*}
CQR_{a}(w) = \frac{w\cdot Cov_{im}}{\sigma_{a}} 
\end{equation*}

where $Cov_{im}$ is the covariance of the $i^{th}$ asset against the entire market, $w$ is a vector of weights for assets in our portfolio and $\sigma_{a}$ is the standard deviation of the collection of assets

\begin{equation*}
CQNS(w;\alpha) = Var(R_{w})-\mathbb{E}[R_{w}]^{2+\alpha}
\end{equation*}

$R_{w}$ is a weighted portfolio and $\alpha \in \mathbb{R}$, which was chosen as to keep an equal weighting i.e. $wi$ = 1/n where n is the number of assets included, and $\alpha$ was kept near 1.

In the case of 40 assets, D-Wave’s 2000 Quantum Annealer performed well. They obtained better results concerning CQNS (achieving stocks with better returns over the specified amount of risk) than the Monte Carlo methods but underperformed the genetic algorithms. In the case of 60 assets, the D-Wave Quantum Annealer was also able to find the ideal portfolio along with the classical methods including simulated annealing. This study was further extended by performing portfolio optimization on 3171 US equity stocks in \cite{cohen2020picking}. The approach here was to first select a few attractive portfolios from the entire universe of portfolios using classical solvers like Monte Carlo, and then use quantum annealing to find the best portfolio. They make use of D-Wave’s Advantage quantum annealer.

Another study using D-Wave’s Advantage quantum annealer\cite{article1} proposed using a classical preprocessing step with a modified QUBO model. The classical preprocessing step includes backtesting a trading strategy and computing the Sharpe ratio and variance on these returns. Then the top 18 asset combinations based on the Sharpe ratio are selected for the QUBO formulation for portfolio optimization. The implementation of the method using quantum annealing showed potential but was outperformed by simulated and digital annealing concerning the Sharpe ratio.

Now let us talk about some gate-based approaches. QAOA \cite{farhi2014quantum} and VQE \cite{Peruzzo_2014, Peruzzo2014} algorithms are found to be good candidates for solving combinatorial optimization problems on the NISQ-era devices. QAOA is well suited for gate model quantum devices. In \cite{Barkoutsos_2020 }, the authors propose a method to improve the results by measurement system by using CVaR(Conditional Value at Risk).

The main objective of QAOA and VQE is to find the optimal solution for the problem by finding the global minimum (or a point close enough to the global minimum) of the energy landscape describing the problem. The energy of the system is encoded in the expected value of the Hamiltonian given by $\langle {\psi(\theta)}| H |{\psi(\theta)}\rangle$. This is termed the objective equation.

QAOA and VQE aim at solving the following equation:
\begin{equation*}
    min_{\theta} \langle {\psi(\theta)} H {\psi(\theta)} \rangle
\end{equation*}

The solution to the above equation is the point of the global minimum. 

A reasonable approximation for the above could be arrived at by using a sample mean. The Hamiltonian for the combinatorial optimization problem is generally framed in such a way that it is diagonalizable and thus there exists a basis state which is the ground state. The eigenvalues lie along the diagonal of the diagonalized Hamiltonian and the minimum of all the eigenvalues corresponds to the ground state. However,  it is difficult to diagonalize a Hamiltonian of a very large dimension and thus an alternative is to perform finite measurements and choose the minimum eigenvalue corresponding to those measurements. The minimum of finite values is not a smooth objective function and thus, CVaR is employed. A measurement system using CVaR as the objective function is especially suitable in such a case.CVaR as the objective function is the expected value of the lower alpha tail of the distribution of X, where X stands for all eigenvalues corresponding to measured states.

The use of $min_{\theta} $CVaR$_{\alpha}(X(\theta))$ (instead of X($\theta$ ))as the objective function, both smoothens the objective function and improves the best-measured outcome. This approach provides a faster convergence to the solution as compared to the use of sample mean and hence provides faster results.

In \cite{Mugel_2021} the authors developed a classical-quantum dynamic portfolio optimization algorithm with minimal holding period constraint to avail the tax benefits, enforced using post-selection. The authors make use of a 50-asset portfolio use case. If one uses brute force methods and successively checks whether the given investment trajectory complies with the minimum holding period constraint, the process becomes computationally prohibitive given the exponentially growing number of investment trajectories. Under post-selection, the investment trajectories are efficiently ruled out, in a tree-like pattern if they fail to meet the constraints. Following the approach, the problem is formulated into a QUBO and is solved using D-Wave’s 2000 Quantum Annealer. Post-selection proved to be faster than the Brute-force classical search approach, which faces issues due to the exponential growth of the required number of qubits as the  number of assets increases.

Apart from QAOA and quantum annealing, we also have an adiabatic quantum optimization algorithm. In order to enable the implementation on a gate model quantum computer, a digitized version of adiabatic quantum computing was introduced \cite{Barends_2016}.In \cite{Hegade_2022}, digitized counter adiabatic quantum computing (DCQC) and digitized counter adiabatic QAOA (DC-QAOA) were studied. When solving a portfolio optimization problem using the adiabatic theorem, the optimal portfolio is represented by the ground state of the problem Hamiltonian. The aim is to increase the success probability of finding this ground state also accounting for the limited coherence time and device noise, which is achieved by using Approximate counter adiabatic-driving(CD)\cite{Sels_2017} terms. 

In adiabatic systems, the solution is obtained by allowing the initial Hamiltonian to evolve to a Hamiltonian whose ground state overlaps the ground state of the problem Hamiltonian. The system needs to be given enough time to evolve to the desired Hamiltonian. However, in the current devices, this becomes difficult due to limited coherence time and device noise. When an adiabatic system is forced to evolve very fast it results in non-adiabatic transitions between the eigenstates. These transitions, in turn, hinder the results. Thus a solution was proposed \cite{Torrontegui_2013,Gu_ry_Odelin_2019}, where one introduces an additional term called counter diabatic-driving term (CD term) so that these excitations are compensated, and the resulting evolution will be quasi adiabatic. In Digitized-counterdiabatic quantum approximate optimization algorithm (DC-QAOA), counterdiabatic (CD) driving is utilized to introduce an additional unitary $U_{D}(\alpha)$, known as the CD term. The paper shows instances where DC-QAOA outperformed QAOA with respect to success probabilities in finding the ground state.

QAOA requires the cost function to be modified to enforce the constraints, using a penalty term. However, this approach is less effective as it needs to search for the best solution in a very large space of solutions.
In \cite{wang2022classicallyboosted}, the authors use a hybrid algorithm, which uses classical methods to find an approximate solution called seed and then uses continuous-time-quantum- walk algorithm \cite{Marsh2019, 
PhysRevResearch.2.023302}. This approach reduces the search area to a smaller subspace. The empirical evaluation of a constrained and an unconstrained problem, showed that the proposed algorithm outperformed the classical alternatives.

Financial Index tracking is an important problem, where we require a small subset of assets to describe the behavior of many assets. Cardinality Constraints help to limit the assets in the portfolio, however, they also lend the optimization problem a non-convex nature. In \cite{palmer2022financial}, the authors use discretized portfolio optimization - which can't be implemented easily using classical devices but can be efficiently implemented using quantum computing- to directly implement cardinality constraints in a single optimization procedure. This approach is successful in generating smaller portfolios that were able to closely track the returns of the NASDAQ 100 and S\&P 500 indexes. The paper also explored enhanced index tracking and were able to construct tracking portfolios that maintained high degree of correlation with the target indexes.

In \cite{rubiogarcía2022portfolio}, the authors worked on finding the optimal portfolio when discretized convex and non-convex cost functions are considered using the integer version of the simulated annealing method. The studies included a multi-period portfolio optimization problem where the non-convexity was introduced from the fixed transactional costs. A multi-period portfolio optimization problem is one where assets are traded at each rebalancing step and thus incur transactional costs. Ignoring the fixed transactional cost while formulating the cost function results in poorer performance than when fixed transactional costs are incorporated in the version of the discrete simulated annealing algorithm used in \cite{rubiogarcía2022portfolio}.

At times investors may be required to make decisions without complete access to information, or when the information is revealed step by step. Such problems are a part of the sub-family of optimization problems called online- optimization problems. In \cite{lim2023quantum}, the authors developed a quantum sampling version of an existing classical online portfolio optimization problem \cite{Helmbold1998OnLinePS}, which provides a quadratic speedup concerning the number of assets in the portfolio and has a transaction cost independent of the number of assets.

The success of quantum computers to solve Portfolio optimization problems using QAOA-based mean-variance portfolio optimization(QAOA-MVPO) was benchmarked in \cite{baker2022wasserstein}. The study compared the performance of quantum simulators(dense state vector simulation and stochastic shot-based simulation) and real devices provided by IBM (superconducting qubits), Rigetti (superconducting qubits), and IonQ (trapped ion qubits). Solution quality was determined by the use of normalized and complementary Wasserstein distance, $\eta$, which allows the QAOA to be intuitively viewed as the transporter of probability. The researchers emphasized on the need for application-specific benchmarking instead of general benchmarking for application performance.

Benchmarking the various versions of QAOA with respect to its suitability to the current hardware is essential to understand which version is most suitable to the developer's needs. In \cite{Brandhofer2022}, the authors tackle this problem and present a detailed analysis of the performance of different versions of QAOA. They also study the influence of statistical sampling
errors and gate and readout errors. The performance of QAOA in finding the solution differs for different instances- when different assets comprise your portfolio.The authors define a criterion for distinguishing between ‘easy’ and ‘hard’ instances of the portfolio optimization problem, by suggesting that instances consisting of stocks with broadly distributed correlations and returns are easier to optimize than those with more similar correlations. This can be attributed to the fact that since the variances of the correlation and returns are high, which causes the variance of the objective function values also increases. This increase in the variance of the objective function causes the energy landscape to be quite distinctive i.e the portfolios are quite different from each other. The problem optimizes a portfolio of assets from the German index DAX.

In \cite{Liu_2022}, the authors introduced a new approach of Layer-VQE, a variation of VQE, for combinatorial optimization problems. The idea here is to start with one layer of parameterized rotations, and then increment the size of the ansatz, by adding entangling gates and other parameterized rotations. The new layer is added before the convergence is reached and this helps to avoid the problem of local minima. This is one very critical requirement to reach the optimal solution - the optimal solution in our portfolio optimization set-up is the optimal portfolio. L-VQE was numerically shown to have much better performance compared to QAOA w.r.t gate count, the gate count of QAOA increases quadratically while that of LVQE increases linearly, and VQE w.r.t handling of finite sampling error and the approximation ratio (ratio used to check the quality of the result, the higher, the better), which increases with each layer of L-VQE ansatz, and decreases with each layer of VQE ansatz. The many-body terms in the Hamiltonian also make QAOA implementation harder. 

In \cite{certo2022comparing}, the researchers include constraints that account for the fundamentals of the companies, using the current ratio and also handle the allocation of assets in each industry.

We also need to ensure that our problem correctly handles our constraints, preferably arbitrary constraints, in order to ensure compliance with the ever-changing dynamics of the market and regulations and give us an optimal solution. In  \cite{herman2023portfolio}, the authors tackle the problem of portfolio optimization, with:

Case 1: an inequality constraint on the total size of the portfolio i.e.
\begin{equation*}
\sum_{j}x_{j} \leq C 
\end{equation*}

Case 2:  As an addition to the above inequality constraint on portfolio size, the authors include a constraint on the total expected return i.e.
\begin{equation*}
    \sum_{j}\mu_{j}x_{j} \geq R 
\end{equation*}

Constraints can be introduced in the problem by either introducing a penalty term or by restricting the evolution of the solution system to an in-constraint subspace. In QAOA, the mixing operator ensures that the evolution occurs in the constraint following subspace, but designing an efficient mixing operator is a difficult process, with additional requirement for efficient Trotterization (the process which breaks the evolution into smaller components) \cite{Fuchs_2022}. In \cite{herman2023portfolio}, enforcing the constraints in the evolution is achieved by the authors by using Quantum Zeno Dynamics via repeated projective measurements which restricts the evolution of the system within an in-constraint subspace. The repeated projective measurements are added to the mixing operator in the QAOA algorithm. This approaches allows both equality and in-equality constraints can be enforced and this is achieved by constructing a quantum oracle.

For QAOA, according to the scaling rule derived by the authors: the number of measurements grows linearly with number of layers in QAOA and quadratically with the number of qubits.

 Upon comparing the performances between Zeno dynamics enhanced QAOA and QAOA using a penalty component, we see that QAOA with zeno dynamics was consistently able to achieve higher approximation ratios, than the penalty approach. Also, the penalty approach requires independent tuning of the penalty factor. When Zeno dynamics was employed in L-VQE, the researchers were able to achieve high approximation ratios and high in-constraint probability. Also, multi-constraints could be employed in L-VQE and not in QAOA due to extremely expensive tuning. For both QAOA and L-VQE the in-constraint probability can be increased using more measurements. 

The application of quantum computers in forex management is discussed in \cite{veselý2022application}. The authors focus on risk management, using an algorithm \cite{Woerner_2019} based on quantum Monte Carlo, and portfolio construction using QUBO with QAOA approach and HHL approach using IBM Quantum processors.They use a small - 5 asset portfolio optimization problem for their study. For portfolio construction using QUBO-QAOA, it was observed that the simulators performed at par with the classical devices as they selected the most efficient portfolios.The quantum devices returned the correct solution as well.In the real quantum devices the difference in performance was with respect to the number of iterations. The number of iterations is greatly influenced by the quantum volume \cite{enwiki:1148963114}. However for portfolio optimization using HHL, the quantum devices suffered from failure due to decoherence and  the inability of the current devices to handle negative eigenvalues in matrices. The risk management algorithm failed to calculate the risk parameters mostly probably due to the difficulty to differentiate between the low percentiles. This stems from the extremely small difference between the angles of the rotational gates. 

HHL algorithms suffer from drawbacks like the ones expressed above due to the limitations of the current NISQ era devices. The researchers in\cite{yalovetzky2023nisqhhl}, thus proposed a NISQ- HHL Algorithm that was used to solve small-size (6 and 14 assets) mean-variance portfolio optimization problems. This work was considered as an enhancement of the works in \cite{Lee_2019}, as in \cite{yalovetzky2023nisqhhl} they replaced the standard Quantum Phase Estimation(QPE) with Quantum Conditional Logic(QCL) enhanced QPE for eigenvalue estimation which helped to reduce the ancillary qubit count, the SWAP gate count and need for qubit connectivity. They also used additional features like mid-circuit measurement and qubit reset and reuse. The trapped ion Quantinuum system model H1 was used due to its support for QCL, mid-circuit measurements, and qubit reset and reuse. QCL-QPE achieved higher fidelity than standard QPE in their studies.

In \cite{campos2022quantum}, the authors developed an open software solution that used the Quantum Metropolis Hasting algorithm to provide a solution to optimization problems. The Quantum-Metropolis Hasting algorithm\cite{Szegedy2004}, achieves a speed up over its classical counterpart by employing quantum walks to reduce the gap between the eigenvalues which leads to shorter mixing times and reaching the minimum energy state faster. They test the results from the software for the N-Queen problem, an NP complexity search problem used as a benchmark for new Artificial Intelligence algorithms, and find that as the problem scales the Quantum Metropolis Hasting algorithms perform better than the classical Metropolis Hasting algorithm, w.r.t time-to-solutions, i.e the ability to reach the solution with least amount of time.

Apart from quantum strategies, there also exists another category of strategies that makes use of quantum-inspired algorithms. One such quantum-inspired strategy is discussed in the following segment. The researchers employed  global-best guided quantum-inspired tabu search with a self-adaptive strategy and quantum-NOT gate (ANGQTS) for portfolio optimization as it has better searchability than the traditional approach in \cite{9586057} on the US stocks in DOW JONES 30. This also allows flexible Fund allocation in a high-solution space. Global-Best Guided Quantum-Inspired Tabu Search with Quantum-NOT Gate GNQTS is a variant of quantum-inspired tabu search (QTS) that aims at achieving the global-best solution and also makes use of a quantum-NOT gate. The strategy to use Quantum NOT allows them to skip the problem of the local optimal solution. The  addition of the self-adaptive mechanism allows it to handle more complex solution spaces making it ANGQTS. The statistical testing proves that ANGQTS achieves significant improvement over GNQTS on weighted allocation portfolio optimization. The paper\cite{9945589} employs a very similar strategy on the Singaporean stocks.

\begin{table}[h!]
\centering
\caption{Works discussed above can be categorized w.r.t the quantum computation models they are based in.}
\label{table:1}
\begin{tabular}{ |p{3.5cm}||p{7cm}|  }
 \hline
 Quantum Computing Models & Related work surveyed\\
 \hline \hline

 Quantum Annealing-based& \cite{Elsokkary2017FinancialPM} \cite{Venturelli2019} \cite{mattesi2023financial} \cite{cohen2020portfolio} \cite{cohen2020portfolio2} \cite{cohen2020picking} \cite{article1}
\cite{palmer2022financial}\\
\hline

 Quantum Adiabatic-based& \cite{Hegade_2022}\\
 \hline

 Gate-based    &\cite{Barkoutsos_2020 } \cite{wang2022classicallyboosted} \cite{Liu_2022} \cite{herman2023portfolio} \cite{veselý2022application} \cite{yalovetzky2023nisqhhl} \cite{lim2023quantum}\\
 
 \hline
\end{tabular}
\end{table}

\begin{table}[h!]
\centering
\caption{The Gate-based model works have also been classified w.r.t the algorithms they use.}
\label{table:2}
\begin{tabular}{ |p{3.5cm}||p{7cm}|  }
\hline
Algorithms & Related work surveyed\\
\hline \hline
 
QAOA &\cite{Barkoutsos_2020}\cite{wang2022classicallyboosted}\cite{herman2023portfolio}\cite{veselý2022application};       
Benchmarking QAOA:\cite{baker2022wasserstein} \cite{Brandhofer2022}\\
\hline

VQE&\cite{Barkoutsos_2020 } \cite{Liu_2022}\\
 \hline

HHL&\cite{veselý2022application} \cite{yalovetzky2023nisqhhl}\\
 
\hline
\end{tabular}
\end{table}

\begin{table}[h!]
\centering
\caption{Implentation of Constraints have been discussed in the following works.}
\label{table:3}
\begin{tabular}{ |p{7cm}| }
\hline
Works on Constraint Implementations \\
\hline \hline
\cite{Mugel_2021} 
\cite{palmer2022financial} \cite{rubiogarcía2022portfolio} \cite{certo2022comparing} \cite{herman2023portfolio}\\
 
\hline

\end{tabular}
\end{table}

\begin{table}[h!]
\centering
\caption{Table 4 provides a snapshot of the discussion in \ref{Rec_Dev}.}
\label{table:4}
\begin{tabular}{ |p{2.7cm}||p{9.5cm}|  }
\hline
Work Surveyed& Contribution\\
\hline \hline

\cite{Elsokkary2017FinancialPM}(2017)& Portfolio Optimization problem for stocks from the Abu Dhabi Securities Exchange formulated as a QUBO, solved using D-Wave’s simulator\\
\hline

\cite{Venturelli2019}(2019)& Quantum-classical hybrid solution to the Markowitz portfolio optimization problem using the D-Wave Quantum Annealer where the quantum annealer is seeded with a solution candidate found by a greedy local search (classical component) and then a reverse quantum annealing protocol (quantum component) is performed\\
\hline

\end{tabular}
\end{table}
\FloatBarrier

\begin{table}[h!]
\Centering
\begin{tabular}{ |p{2.7cm}||p{9.5cm}|  }
\hline

\cite{cohen2020portfolio}(2020)& Portfolio optimization of 40 US stocks respectively using D-Wave’s Quantum Annealer. Developed CQR(Chicago Quantum Ratio) and CQNS (Chicago Quantum Net Score) as an improvement over the Sharpe Ratio with respect to ease of handling for the annealer\\
\hline

\cite{cohen2020portfolio2}(2020)& Extension of the research in \cite{cohen2020portfolio}; portfolio optimization of 60 US stocks respectively using D- Wave's Quantum Annealer\\
\hline

\cite{cohen2020picking}(2020)& Extension of the research in \cite{cohen2020portfolio} and \cite{cohen2020portfolio2}. Portfolio optimization on 3171 US equity stocks\\
\hline

\cite{Barkoutsos_2020 }(2020) & Proposed a method to improve the results by measurement system by using CVaR(Conditional Value at Risk)\\
\hline

\cite{9586057} (2021) & A quantum-inspired algorithm; employed  global-best guided quantum-inspired tabu search with a self-adaptive strategy and quantum-NOT gate (ANGQTS) for portfolio optimization\\
\hline

\cite{yalovetzky2023nisqhhl}(2021) & Proposed a NISQ- HHL Algorithm that was used to solve small-size mean-variance portfolio optimization problem\\
\hline

\cite{Liu_2022}( 2022)& Introduced a new approach of Layer-VQE, a variation of VQE, for combinatorial optimization problems\\
\hline

\cite{baker2022wasserstein}(2022)& Benchmarked the success of quantum computers to solve Portfolio optimization problems using QAOA-based mean-variance portfolio optimization(QAOA-MVPO)\\
\hline

\cite{wang2022classicallyboosted}(2022)& Used a hybrid algorithm, where classical methods were used to find an approximate solution called seed, and then a continuous-time-quantum- walk algorithm was used\\

\hline

\cite{certo2022comparing}(2022)& Included constraints that portray the fundamentals of the companies\\
\hline

\cite{veselý2022application} (2022)& Discusses the application of quantum computers in forex management by comparing the performance of QAOA and HHL for the same\\
\hline

\cite{campos2022quantum}(2022) & Developed an open software solution that used the Quantum Metropolis Hasting algorithm to provide a solution to optimization problems\\
\hline

\cite{lim2023quantum}(2022)& Developed a quantum sampling version of an existing classical online portfolio optimization problem\\
\hline

\cite{palmer2022financial}(2022)& Tackled the problem of Financial Index Tracking by using discretized portfolio optimization to directly implement cardinality constraints in a single optimization procedure\\
\hline

\cite{herman2023portfolio}(2022) & Ensures that the problem correctly handles arbitrary constraints using Quantum Zeno Dynamics via repeated projective measurements\\
\hline

\cite{Mugel_2021}(2022)& Developed a classical-quantum dynamic portfolio optimization algorithm with minimal holding period constraint to avail the tax benefits, enforced using post-selection\\
\hline
\end{tabular}
\end{table}
\FloatBarrier

\begin{table}[h!]
\Centering
\begin{tabular}{ |p{2.7cm}||p{9.5cm}|  }
\hline

\cite{rubiogarcía2022portfolio} (2022)& Worked on finding the optimal portfolio when discretized convex and non-convex cost functions(introduced via fixed transactional cost) are considered using the integer version of the simulated annealing method\\
\hline

\cite{article1}(2022)& Used D-Wave’s Annealer and proposed using a classical preprocessing step with a modified QUBO model for solving portfolio optimization problems\\
\hline

\cite{Brandhofer2022} (2022)& Benchmarked the various versions of QAOA concerning its suitability to the current hardware\\
\hline

\cite{Hegade_2022}(2022)& Studied digitized counter adiabatic quantum computing (DCQC) and digitized counter adiabatic QAOA (DC-QAOA)for portfolio optimization\\
\hline

\cite{mattesi2023financial} (2023) & Proposed an improvement in the QUBO formulations of \cite{Venturelli2019} allowing the investor to decide the optimal fund allocation in each asset\\
\hline

\end{tabular}
\end{table}
\FloatBarrier

\justify

\section{Fraud Detection and Credit Scoring}
\label{Fraud_sec} 

Fraud detection is a critical function for banks to maintain trust with their customers and protect their assets. However, traditional methods of fraud detection rely on rules-based systems and statistical analysis, which can be limited in their ability to detect sophisticated fraud schemes.

Quantum computing can potentially provide significant advantages in fraud detection by processing vast amounts of data and detecting patterns that may be difficult or impossible to detect with classical computers. Specifically, quantum computing could be used to analyze data from various sources such as transaction records, social media, and other public information to identify fraudulent activity.
Moreover, quantum computing can also help banks in their cybersecurity efforts by providing a stronger encryption for sensitive data, ensuring that data breaches and other types of cyberattacks do not result in financial loss.
Overall, quantum applications for fraud detection at banks is still in its early stages, and significant research is needed to develop the algorithms and infrastructure required for practical implementations. However, the potential benefits of quantum computing in this area could make it a vital technology for the banking industry in the future.

The paper \cite{Grossi_2022} describes the first-ever end-to-end application of a Quantum Support Vector Machine (QSVM) algorithm in the financial payment industry using IBM Safer Payments and IBM Quantum Computers via Qiskit software. The study uses actual card payment data to compare the performance of state-of-the-art quantum machine learning algorithms with the classical approach. The researchers also explore a new method to search for the best features using QSVM's feature map characteristics. Key performance indicators such as accuracy, recall, and false positive rate are compared for the classical machine learning algorithms (Random Forest, XGBoost), quantum-based machine learning algorithms using QSVM, and human expertise (rule decisions). Additionally, the paper explores a hybrid classical-quantum approach that uses an ensemble model combining classical and quantum algorithms to improve fraud prevention decisions.

Three approaches are compared on the same dataset: 
\begin{enumerate}
\item Domain expert created decision rules-based model (no machine learning) 
\item State-of-the-art type AI/ML using boosted trees (Random Forest, XGBoost) 
\item Quantum Support Vector Machine (QSVM) type model.
\end{enumerate}

The authors use data comprising of 2.4 million payment transactions. Each transaction is flagged as fraud or non-fraud, with a total of 3K transactions marked as fraudulent. There are 12 features from transactional data, 2 features are from demographic data and the remaining features are generated through discovery techniques. Since the data is highly imbalanced, the authors tried some types of under-sampling techniques. 

Among all possible quantum approaches, \cite{Grossi_2022} focuses on QSVM. The motivation for this work is to leverage the QSVM approach in two parts in order to optimize the fraud detection system. The first is to determine which of the many features available should be selected to reduce the dimensionality of the data set for running the experiment on a quantum system. The second is to derive the fraud KPIs from a quantum machine learning model. Using a Quantum approach also requires restricting to a few important features so that number of qubits required are not too many and to reduce the number of data points. Using undersampling techniques to scale down data is an important prerequisite. All data values are also normalized using MinMaxScaler package as a more convenient choice for quantum processing. 

For feature selection, instead of using standard classical approaches of feature importance, PCA or Factorial Analysis of Mixed Data (FAMD) method, the authors developed a quantum algorithm that would allow use of a quantum feature map and quantum kernels to determine best features. 

The authors employed the Qiskit framework, which includes a QuantumKernel class and a ZZFeatureMap, in their implementation. The ZZFeatureMap was utilized to map each data point to a Quantum State, and the resulting inner products of these states were utilized to generate the Kernel matrix. Their methodology was inspired by the Feed Forward Feature Selection (FFFS) approach, which is based on statistical metrics such as AUC or Accuracy. By using this approach, the authors were able to iteratively select an increasing number of features in the problem, beginning with only three out of a total of 69. 

The results with Random Forest, XGBoost and QSVM on a Balanced dataset after undersampling show that the AUC on the test data is more or less similar for all the approaches with an acceptable difference of about 0.01, while the accuracy on the test data is 2 percent higher with QSVM than other approaches.

The mixed quantum-classical approach: To improve classification performance, the authors combine the strengths of both quantum and classical algorithms. To do this, they identified transactions or data points where the quantum and classical algorithms have different classifications. They then trained a metaclassifier on these "disagreed" data points to predict which classification is correct. They trained both the quantum and classical algorithms on the training data set, and noted any transactions where the two algorithms disagreed. These transactions formed a smaller dataset, on which they trained a metaclassifier that could use any feature from the original dataset. Due to the limited number of disagreed data points, a simple metaclassifier worked best.

To summarize, the authors found that quantum classifiers can detect patterns that classical algorithms struggle with, and a mixed quantum-classical ensemble can improve the final model. The results are obtained on a simulated quantum computer, and future work will explore real hardware implementation. Pre-processing the data is crucial before moving to the quantum part.

Fraudulent transactions could be considered as anomalous events since it is expected that the non-fraud or genuine transactions are always more dominant. The work of \cite{kyriienko2022unsupervised} focuses on a Quantum approach to detect frauds via anomaly detection. Anomaly detection is an unsupervised approach since it does not require pre-defined labels for transactions as fraud or non-fraud. 

The approach involves using IQP or Instantaneous Quantum Polynomial as a feature map. This feature map would map the original data into a high dimensional data. Further, a quantum kernel approach or QSVM is used. Though SVM or QSVM are generally considered as supervised learning approaches, the approach mentioned in this paper is a Quantum analogue of the classical one-class SVM, which is a popular classical approach for anomaly detection. The idea here is that inner products of resultant quantum states used in the kernel matrix would adequately show more distance for data points in different classes vs data points in the same class. 

For feature enrichments/enhancements or feature engineering prior to embedding data with IQP, the authors tried some scaling strategies for all the features, but that did not lead to any performance improvement. 

For defining and simulating the quantum circuits, the authors used Pennylane library for Python, with the JAX interface that allows for compiling the circuits with XLA (Accelerated Linear Algebra or XLA is a compiler that uses JIT compilation techniques with the purpose of speeding up extensive ML workloads). This results in fast, parallelized operation across CPU and GPU resources. Additionally, they implemented batched Gram matrix evaluation with kernel evaluations streamlined over different values. 

At small number of features the authors observed increase of average precision up to 0.9 values for five-to-ten features. They noted that the classical kernel model performance deteriorates as they tried beyond 10 features, while the quantum kernel model performance does not, and even reaches new highs around 17 features/qubits. At N = 20, they saw a clear separation of performance between quantum and classical (RBF) kernels. This suggests that quantum kernels avoid overfitting. This is a positive sign showing expressivity/learning advantage for certain tasks performed on classical datasets. 

In  \cite{tapia2022fraud}, the authors present a novel technique for performing classification, with a specific focus on fraud detection as a real-world application, utilizing a single qubit. This approach is particularly advantageous in the current NISQ era, where quantum hardware with a large number of qubits is not yet widely available, and the systems that do exist with a higher number of qubits are susceptible to noise and errors. 

The approach takes inspiration from the concept of data-reloading proposed in \cite{P_rez_Salinas_2020} which allows to encode mathematical functions in the degrees of freedom of a series of gates applied to a single-qubit state. This strategy is inspired in the concept of classical neural networks, where the computations are performed by neurons organized in interconnected layers. In this quantum analogy, the unitary rotations can be seen to correspond to neurons, and form processing units that can be replicated to create layers. Each subsequent neuron “re-uploads” the classical input data, and is able to capture a particular feature of the distribution. 

Additionally, in this paper \cite{P_rez_Salinas_2021}, it is shown that a quantum neural network based on a single-qubit circuit can approximate any bounded complex function by storing its information in the degrees of freedom of a series of quantum gates.

In the case of the single-qubit QNN classification, the feature map and ansatz both are parametrized as arbitrary single-qubit rotations. Each combination of the data encoding single qubit rotation and parameterized rotation can be considered as analogous to a layer in the classical neural network architecture. 

\begin{equation}
   U(\vv{o{}},\vv{x}) \equiv U(\vv{o{}_{N}}) U(\vv{x}) ... U(\vv{o{}_{1}}) U(\vv{x})   
\end{equation}

Given that each layer is composed of two unitaries, the total depth of a circuit with N layers will be 2N. According to the UAT, the more layers there are, the more representation capabilities will be present in the circuit. However, the more layers in a circuit the more time it will take to run. This may affect negatively the quality of the results due to the limited coherence times in current quantum processing units. Therefore, in \cite{P_rez_Salinas_2020}, the authors came up with the unitary of the below form (the symbol o in the formula below indicates Hadamard product).

\begin{equation*}
   L(i) = U(\vv{\theta_{i}} + \vv{\omega_{i}} o \vv{x} )
\end{equation*}

The issue with above is that it leads to computational complexities for higher depths. Therefore, in \cite{P_rez_Salinas_2021}, below was proposed, which is referred to as a fundamental UAT (a single qubit Universal Approximant) gate.

\begin{equation}
   U^{U AT}(\vv{x},\vv{w},\alpha,\varphi) = R_{y}(2\varphi)R_{z}(2\vv{\omega}\cdot \vv{x} + 2\alpha),    
   ({\vv{\omega},\alpha,\varphi}) \in ({R_{m}},R,R)
\end{equation}

Once the feature map and ansatz are defined for a target variational quantum circuit, it can be trained following the typical hybrid procedure. The input data is loaded into the network with an initial set of arbitrary parameter values. The gates are applied and followed by a measurement operation at the end. The result of this measurement is fed into a specific cost function that is used to guide a classical optimizer to find the next set of parameters. This process is performed iteratively until the optimizer reaches the minimum cost.

A quantum measurement strategy is incorporated to find the optimal way to associate the outputs from the quantum observations to the target classes. A standard approach is to use threshold(s) to map outputs from measurements to a target class. For e.g., for binary classification, P(0) less than or equal to the threshold would imply that the datum could be mapped to class 0, else it would be mapped to class 1. 

A fidelity-based loss function is used and the classical optimizer used is LBFGS:

\begin{equation*}
   {X^{2}}_f(\vv{\theta},\vv{\omega}) = \sum_{\mu=1}^{M}(1- {|\langle\psi^{\mu}_{s} | \psi(\vv{\theta},\vv{\omega},\vv{x}_{\mu})\rangle|}^2)
\end{equation*}

where $\mu$ indicates the data point and $|\psi^{\mu}_{s}\rangle$ is the correct label state.

The authors tried the approaches on both a toy data (circles data) and on the real-world credit card data from Kaggle. The original approach to use data encoding unitary and parameterized unitary separately gave better accuracy on both training and test data, as compared to Compressed unitary and UAT

In order to boost the training process with UAT, an initial data loading layer might be of help. An initial hypothesis is that a data preparation step using a Hadamard gate would help take advantage of the state of superposition for facilitating the parameter search. Following this train of thought, it could be even more beneficial to perform the data preparation state with a generic parametrized unitary gate, whose parameters are optimized jointly with those of the ansatz. On the toy data, initial loading with parameterized unitary gate U, gave better training and test accuracy than having no initial loading or having a Hadamard gate as the initial loading.

The authors also experimented with different numbers of layers. Using 4 layers with UAT combined with initial loading with an arbitrary unitary rotation gave the most optimal results for both training and test accuracies.

They also did a comparison of results for UAT with and without the initial data loading with a parameterized unitary gate. With initial loading, the test accuracy was 2 per cent more than the test accuracy without initial loading. The train accuracy was 2 per cent higher without initial loading. This suggests that the initial loading can also combat potential overfitting. 

For the real-world data, the authors also performed PCA for dimensionality reduction and data sampling to reduce data samplings and combat class imbalance. The algorithm results for UAT+U method are presented in terms of true positives (tp), true negatives (tn), false positives (fp), false negatives (fn), accuracy (Acc.), precision and recall. 2 layers gave significantly better results than a single layer and slightly better results than 4 layers.

The overall benchmarking results showed that 2 layers of UAT along with initial loading gave better results than the best classical approach with 2 layers. 

While these findings are encouraging, they should not be regarded as definitive evidence of superiority. In such situations, classical machine learning maintains its dominance because it can handle substantial data sets within the framework of deep neural networks. Nonetheless, if a single qubit can effectively learn the pertinent connections between data and corresponding labels, upcoming structures incorporating multiple qubits could refine certain quantum characteristics, such as entanglement, which may potentially confer an advantage.

\justify
\section{Monte Carlo Methods in Finance}
\label{MonteCarlo_sec}

The application of Monte Carlo methods to finance started in 2018 with the seminal paper of Rebentrost et al \cite{Rebentrost_2018}. In their paper, they applied earlier theoretical work about the use of quantum methods to speed up Monte Carlo simulations, especially \cite{Brassard_2000} and \cite{Montanaro_2015}. For this review of Monte Carlo methods in finance we used \cite{Rebentrost_2018} as a starting point.

In the following two subsections, we will cover the main fields in finance, where Monte Carlo methods are used: derivative pricing and risk calculation.
A survey covering these two topics and thereby concentrating on the details of the calculations can be found in \cite{Gomez_2022}.

\subsection{Derivative Pricing}

\subsubsection{The basis: Quantum Computational finance}
The application of quantum computing to solve simulation bases problems in finance started with the paper  \cite{Rebentrost_2018}. Here, one of the classical problems in finance -the pricing of derivative contracts - is considered:
\begin{equation}\label{eqPrice}
\Pi = e^{-r T}  E_{P} [f(S_T)].
\end{equation}
Here, $r$ is the risk-free interest rate, $T$ the run-time of the contract, $E_P$ the expectation value under the risk-free measure $P$, and $f(S_T)$ the pay-off function of the derivative at time $T$, depending on the corresponding underlying. 
As an example, the case of a plain-vanilla European call (or put) option is considered: A European call (put) option gives the owner of the contract the right to buy (to sell) the underlying option for a predefined price at a certain time event in the future.
Here, eq. (\ref{eqPrice}) can be solved analytically \cite{Hull1993}. However, when the pay-off of a derivative becomes more complicated this is usually not the case anymore and people switch to e.g. Monte-Carlo simulation to calculate the price $\Pi$ in (\ref{eqPrice}). 
When the pay-off $f(S_T)$ has bounded  variance $\lambda^2$ a constant success probability requires 
\begin{equation}\label{eqStepsClassical}
k= {\cal O} (\frac{\lambda^2}{\epsilon^2})
\end{equation} samples to estimate the expected value up to an error $\epsilon$.

Using the available exact result for European call options, \cite{Rebentrost_2018} shows that a quantum-enhanced Monte-Carlo method requires only $k= {\cal O} (\frac{\lambda^2}{\epsilon})$ steps, thereby resulting in an exponential speedup.

The idea of using a quantum algorithm to do a Monte Carlo simulation can be described as follows:
Consider a quantum state
\begin{equation}
\ket{\psi}_n = \sum_{i=0}^{N-1} \sqrt{p_i} \ket{i}_n, 
\end{equation}
with $p_i \in [0, 1]$, $\sum_{i=0}^{N-1} p_i = 1$, and $N = 2^n$. $p_i$ is the probability to measure the state $\ket{i}_n$. 
This state $\ket{i}_n$ is one of the $N$ possible realizations of a bounded discrete random variable $X$.

Now consider a function $f: \{0, ..., N-1\} \rightarrow [0, 1]$ which models the pay-off of the derivative and a corresponding operator 
\begin{equation}
{\cal A}: \ket{i}_n\ket{0} \mapsto \ket{i}_n \left( \sqrt{1 - f(i)}\ket{0} + \sqrt{f(i)} \ket{1} \right),
\end{equation}
for all $i \in \{0, ..., N-1\}$, acting on an ancilla qubit.
Applying ${\cal A}$ to $\ket{\psi}_n\ket{0}$ leads to the state
\begin{equation}
\sum_{i=0}^{N-1} \sqrt{1 - f(i)} \sqrt{p_i} \ket{i}_n\ket{0} 
+ \sum_{i=0}^{N-1} \sqrt{f(i)} \sqrt{p_i} \ket{i}_n\ket{1}.
\label{Basic_eq}
\end{equation}

Finally defining a suitable unitary operator $Q$, c.f. \cite{Rebentrost_2018} and applying $Q^k, k=0, 2^{n-1}$ 
in a controlled way to the circuit leads to an amplification of the desired amplitude. With an inverse Quantum Fourier Transformation, it is possible to extract the phase of this state which is then measured as a bit-string.

The structure of the proposed algorithms consisting an amplitude amplification and phase estimation is shown in the following circuit:

\begin{figure}[ht]
	\centering
		\includegraphics[width=.45\textwidth]{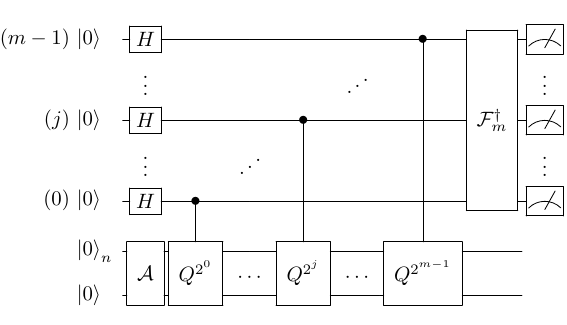}
	\caption{Full quantum circuit for the Monte-Carlo algorithm, c.f. \cite{Brassard_2000}. $H$ is the Hadamard gate and $\mathcal{F}_m^{\dagger}$ denotes the inverse Quantum Fourier Transform on $m$ qubits.}
	\label{fig:Full_MC_Circuit}
\end{figure}

The results of \cite{Rebentrost_2018} can be summarized as following:
\begin{enumerate}
    \item Under relatively mild assumption (bounded variance of the random distribution) a quadratic speed up of the quantum algorithms compared to classical Monte Carlo is proven.
    \item In the case of a European call option where the random distribution is normal, the gates to prepare the algorithm are explicitly shown and give a guide to implement it. 
    \item In a numerical study for a small setup the quadratic speed-up of quantum Monte Carlo is demonstrated explicitly, cf. Fig.3 of \cite{Rebentrost_2018}
\end{enumerate}

\subsubsection{Improvements of the basic approach}
The approach for derivative pricing shown in the last section, is very interesting from an application point of view since it offers a quadratic speed-up compared to classical Monte Carlo. However, its implementation in a realistic setup is beyond the capability of today's and also near-future quantum hardware. Neither the amount of qubits nor the circuit length is and will be available in the short term. Therefore, soon after the publication of \cite{Rebentrost_2018} the search for modifications began, which looked for alternatives in using quantum amplitude amplification and quantum phase estimation together. In a recent review \cite{Intallura_2023} technical details of the different lines of improvements are demonstrated and discussed.

In \cite{Suzuki_2020} an "amplitude amplification without phase estimation" approach is proposed. Instead of applying all the controlled ${Q^k}$ followed by an inverse Fourier transformation, it is suggested that ${Q^k}$ is applied only once, followed by a direct measurement. Doing this several times with different values of $k$ a likelihood function for the desired target parameter is obtained. This likelihood can then be maximized. In \cite{Aaronson_2020}, it is theoretically examined if and how a quadratic speedup of the Monte Carlo algorithms without Quantum Fourier transformation can be achieved. In a subsequent publication \cite{Tanaka_2020} results of the maximum likelihood method in a realistic environment, i.e. on a real quantum device of IBM, are shown. It is demonstrated, how the different errors affect the theoretical results. A comparable analysis, using theoretical noise models instead of real quantum devices can be found in \cite{Brown_2020}.
 
Following the lines of \cite{Tanaka_2020} in \cite{Uno_2021} a modification of the Grover operator is suggested which proved to be superior in the case of noise existing on physical devices.

On the basis of \cite{Suzuki_2020} in \cite{Plekhanov2022} the so-called (adaptive) variational quantum amplitude estimation is developed. With the help of a variational ansatz, the circuit depth of the usual amplitude estimation algorithm is reduced. The maximum likelihood method is applied for post-processing and it is demonstrated, that classical Monte Carlo is outperformed.

Another approach to circumvent the expensive circuit described above is suggested in \cite{Grinko_2019}. The authors call their method
Iterative QAE (IQAE) and claim that it achieves better results than comparable algorithms. The idea is to iteratively find the optimal power $k$ of the Grover operator ${Q^k}$. An algorithm to achieve this is explicitly provided and numerical results are shown. Furthermore, the method is compared to the maximum likelihood method discussed above. However, a drawback of this method is the fact, that the calculations can not be performed in parallel.

A Monte Carlo algorithm without the Quantum Fourier transformation but with extensive post-selection is also suggested in \cite{Ramos-Calderer_2019}. Here, a unary encoding of the asset value is used. This leads to a reduction of the circuit depth, however, much more qubits are needed compared to the usual approach. It is stated that this approach may be advantageous in the NISQ era.

While in most publications the pricing of a plain-vanilla European option is considered, \cite{Kaneko.2022} extends the framework to pricing options with a local volatility model i.e. in which the volatility of the underlying asset depends on the price and time. The solution is provided using amplitude encoding where the probability distribution of the derivative’s payoff is encoded to the probabilistic
amplitude. Additionally, in a second method sequences of pseudo-random numbers are used to simulate the asset price evolution as in classical Monte Carlo
simulation. The idea of using pseudo-random numbers is further elaborated in \cite{Kaneko.2021}: In case the integrand used in the Monte Carlo calculation has a separable form with respect to contributions from distinct random numbers a combination of a nested quantum amplitude estimation and the use of pseudo-random numbers can speed up the calculation.

In \cite{Stamatopoulos2020} the framework of quantum Monte Carlo is applied to the pricing of different option types: Plain vanilla options, multi-asset options, and path-dependent
options such as barrier options. The paper puts an emphasis on implementation details and demonstrates, how non-trivial payoff functions can be encoded into the Quantum circuit. For the evaluation, the approach of \cite{Suzuki_2020} is used, i.e. without Quantum Fourier transformation and with Maximum Likelihood post-processing. Results are shown when using real hardware of IBM and a sophisticated error reduction scheme is applied. 
In the subsequent paper, \cite{Stamatopoulos2022} the calculation is extended to determine the option Greeks, i.e. the derivatives of the option price with respect to the input parameters. The calculation of these derivatives is an essential part of determining the market risk of financial options. Another approach to calculating derivatives of financial options is discussed in \cite{Miyamoto.2022_1}.

Some of the just mentioned authors published in \cite{Chakrabarti2021} an upper bound on the resources required for valuable quantum advantage in pricing derivatives. 
They considered a special kind of derivative, autocallable, and Target Accrual Redemption
Forward (TARF) derivatives as benchmark use cases. Even after improving the existing algorithm by combining pre-trained
variational circuits with fault-tolerant quantum computing they find that the benchmark use cases require 8000 logical qubits and a circuit depth of 54 million. However, the pricing could be done in just one second. 

Another more complex financial derivative is examined in \cite{Miyamoto.2022}: For the pricing of Bermudan options which can be executed at specific dates it is necessary to model the payoff more thoroughly. It is shown how a combination of  Chebyshev interpolation and quantum amplitude estimation can be used in this case.

In the domain of interest rate derivatives, the LIBOR market model plays an important. How e.g. caps can be priced in the Quantum Monte Carlo scheme with this underlying model is elaborated in \cite{Tang_2022}.

 The existing weaknesses of the full Monte Carlo algorithm, especially in the NISQ era, inspired further work to improve the procedure on a fundamental level. The authors of \cite{Herbert2022} suggest a  method for Monte Carlo integration by expanding the integral as a Fourier series and then estimating each component individually using amplitude estimation. Via a theoretical calculation as well as via numerical results it is asserted, that the full quadratic speedup can be achieved with this method. 

 Another fundamentally different approach is described in \cite{Saha_2022}: In order to do the Quantum Monte Carlo algorithm without any ancilla qubits it is suggested to use intermediate qutrit states with three levels instead of two. 

 In order to circumvent the computational load of the full Monte Carlo approach \cite{Braun2022} propose a method where the quantum amplitude part of the algorithm is performed in parallel. This reduced the circuit depth; however, requires more qubits. 

 Whereas in most analyses it is assumed that the payoff function characterizing the relevant option is positive, this assumption is removed in \cite{Manzano_2023}. Furthermore, in this paper a fully worked-out calculation of different options, using the above-mentioned quantum amplitude estimation schemes is presented.

A crucial point of all Quantum Monte Carlo methods is the loading of the probability distribution into the quantum state. If this is done not carefully enough computational advantages can be destroyed. In \cite{Vazquez2021} an approach is presented that simplifies state preparation. Together with a circuit optimization technique, it can help to reduce the circuit complexity for QAE state preparation significantly. Results are presented for option pricing under a specific stochastic volatility process, the Heston
model, on real hardware. Another approach is described in \cite{Zoufal_2019}: There a Quantum Generative Adversarial Network is used for learning
and loading random distributions into the quantum circuit which can then be exploited with the subsequent Monte Carlo steps.

An extension of the derivative pricing method via Monte Carlo to another class of financial products is the topic of \cite{Tang_2020}: It is described how the tranches of collateralized debt obligations (CDO) can be priced via a Quantum Monte Carlo method. As the underlying algorithm, the iterative QAE of \cite{Grinko_2019} is implemented, and a suitable credit default model, the Gaussian copula, is used for the underlying dynamics.

Finally in \cite{Alcazar_2022} a detailed calculation of credit valuation adjustments (CVA) is presented. This topic is also addressed in \cite{Han_2022}, but from a purely theoretical perspective.

\subsection{Risk calculation}
The application of Quantum Monte Carlo methods to risk calculations in the financial industry started with the paper Quantum Risk Analysis of Woerner and Egger 2019, \cite{Woerner_2019}. They used the full algorithm, i.e. quantum amplitude estimation including the quantum Fourier transformation to calculate the Value-at-Risk of fixed rate treasury bond.

The basis of the algorithm is again eq.  (\ref{Basic_eq}). Applying amplitude estimation directly to approximate the probability of measuring $\ket{1}$ in the last qubit, leads to
$\sum_{i=0}^{N-1} p_i f(i)$ and thus also $\mathbb{E}\left[f(X)\right]$.

For a given confidence level $\alpha \in [0, 1]$, $\text{VaR}_{\alpha}(X)$ can be defined as the smallest value $x \in \{0, ..., N-1\}$ such that $\mathbb{P}[X \leq x] \geq (1 - \alpha)$. In the same way, the Conditional Value-at-Risk (CVAR) - the expectation value of all values exceeding the VaR can be calculated. 

In \cite{Woerner_2019} the simple example (only one asset) is even further reduced, so that the calculation can be done with four qubits. The expected theoretical quadratic speedup is demonstrated and even results on a physical quantum device are shown.

A direct extension of this work to credit risk is presented in \cite{Egger_2020_1}. The conditional independent Gaussian model which is considered in the paper is very simple but it allows a straightforward implementation of a quantum algorithm. 

A credit portfolio risk calculation is also shown in \cite{Miyamoto_2019}. It is demonstrated, that it is possible to reduce the number of qubits while keeping quantum speed up if the calculation is performed similarly to the classical one: To estimate the average of integrand values sampled by a pseudo-random-number
generator (PRNG) implemented on a quantum circuit. However, the reduction of qubits is
a trade-off against an increase in circuit depth.

In \cite{Dri_2022} a method is described to
implement a more realistic and complex risk model for the default probability of each portfolio’s
asset, capable of taking into account multiple systemic risk factors. Additionally, the constraint that the Loss Given Default may only take integer values is removed. The approach is implemented and the corresponding gate structure is presented.

A very important question for potential applications in finance is addressed in \cite{Miyamoto_2022_3}: How to calculate risk contributions in a quantum-enhanced credit risk model. The necessary circuit structure is theoretically elaborated and a speedup compared to a classical approach is derived.

Another extension of the quantum Monte-Carlo approach for risk calculation is presented in \cite{Matsakos_2023}. There scenario generation
into the quantum computation is incorporated by simulating the evolution of risk factors over
time. To achieve this, quantum
circuits that implement stochastic models
for equity (geometric Brownian motion),
interest rate (mean-reversion models), and
credit (structural and reduced-form credit
models) risk factors are assembled. These scenarios build the input for Quantum Monte Carlo simulations to provide
end-to-end examples for both market and
credit risk use cases.

We close this section by referring to \cite{Wilkens_2023}. Here, the different approaches to calculating market and credit risk via Quantum Monte Carlo are reviewed and demonstrated.

\begin{table}[h!]
\centering
\caption{Works discussed above can be categorized w.r.t the algorithms and/or applications.}
\label{table:GH1}
\begin{tabular}{ |p{3.5cm}||p{7cm}|  }
 \hline
 Algorithms and Applications & Related work surveyed\\
 \hline \hline
Basic Monte Carlo Algorithm for Derivative Pricing& 
\cite{Brassard_2000}
\cite{Montanaro_2015}
\cite{Rebentrost_2018}
 \\
\hline
Improvements of Basic Algorithm& 
 \cite{Grinko_2019}
 \cite{Ramos-Calderer_2019}
\cite{Suzuki_2020}
 \cite{Aaronson_2020}
 \cite{Tanaka_2020}
\cite{Intallura_2023}
\cite{Brown_2020}
 \cite{Uno_2021}
 \cite{Plekhanov2022} 
\cite{Herbert2022}
 \cite{Saha_2022}
\cite{Braun2022}
\cite{Vazquez2021}
\cite{Zoufal_2019}
\\
\hline
Application to non-vanilla derivatives &
 \cite{Kaneko.2021}
\cite{Kaneko.2022} 
 \cite{Stamatopoulos2020}
  \cite{Miyamoto.2022_1}
\cite{Chakrabarti2021}
 \cite{Miyamoto.2022}
 \cite{Tang_2020}
 \cite{Tang_2022}
  \cite{Manzano_2023}

\\
\hline

CVA and Risk Calculation& 
\cite{Alcazar_2022}
\cite{Han_2022}
\cite{Woerner_2019}
 \cite{Egger_2020_1}
  \cite{Miyamoto_2019}
   \cite{Miyamoto_2022_3}
\cite{Gomez_2022} 
 \cite{Dri_2022}
  \cite{Matsakos_2023}
  \cite{Wilkens_2023}
 \\
 \hline

 \hline
\end{tabular}
\end{table}

\FloatBarrier

\section{Blockchain in Quantum Finance}
\label{Blockchain_sec}

Quantum computing and blockchain systems are two emerging technologies of recent years that have the potential to revolutionize business models. The latest advances in quantum computing have a significant impact on the computational efficiency of important algorithms which brings together serious security concerns in cryptography related technological systems one of which is the blockchains. In this section, we give a detailed analysis of blockchain based  financial systems in the emerging quantum era. We first discuss the basics of the blockchain concept and its various use cases in the real world followed by specific financial use cases such as cryptocurrencies, smart contracts, digital payment, and exchange systems, NFTs, etc. Then, we will examine the general security and privacy aspects of the financial blockchains with a specific focus on the quantum threats on them. We then discuss the ways to avoid those possible quantum attacks and the difference notion between the concept of quantum-resistant blockchain versus quantum-safe blockchain.  We handle the privacy-preserving quantum-resistant coins apart from the security analysis and finally, we introduce the quantum blockchain. We would like to emphasize that even though there exist several reviews regarding blockchains and quantum technologies, most of them just focus on one aspect of the duo's relation with each other i.e. they either focus on the quantum-resistance of the blockchain or just security aspects of the blockchain without considering the quantum attacks, or just quantum computing and mining.  This review, on the other hand, takes all those aspects regarding the blockchain and quantum technologies as well as the privacy-preserving coins and their usage in finance. In that sense, this review gives a broader perspective of the aforementioned concepts including some security issues that are not mentioned in the other recently published surveys as well.

\subsection{Blockchain Basics} Here we start with a brief conceptual description of the blockchain systems and their basic structure as well as their use cases in various real-world applications. Blockchain is a digital  distributed and decentralized ledger system that is basically a database but instead of a single data source, the data is divided into multiple blocks in various computers which are called peers (nodes). The data structure resembles a public record where all completed transactions are recorded in a sequence of blocks. This sequence expands over time as new blocks are added to it continually. The ruling of this distributed system is held by consensus algorithms.  The evolution of blockchain systems started with the idea of constructing a secure peer-to-peer (P2P) digital payment system without a need for a trusted third party such as banks. User security and ledger consistency have been ensured through the utilization of asymmetric cryptography and distributed consensus algorithms. Key characteristics of blockchain  are:
 
\begin{itemize}
\item 	\textbf{Decentralization:} Blockchain consensus algorithms enable the system to circumvent the need for a trusted third party, along with its associated costs and performance limitations.
\item\textbf{Persistency (Immutability)}: Once transactions are added to the blockchain, it becomes extremely difficult, if not impossible, to remove or undo them.
\item \textbf{Anonymity}: Users are provided with a unique address to access the blockchain, with Bitcoin addresses being generated as a 160-bit hash of the user's public key.
\item \textbf{Auditability}: The use of the UTXO (Unspent Transaction Output) model in blockchain enables straightforward verification and tracking of all transactions, thus ensuring auditability.
\end{itemize}

\subsubsection{Real World Application Areas of Blockchain Technology}
Blockchain finds application in multiple financial services, such as digital assets, online payment, and remittance, as it allows for payment completion without the involvement of banks or central authorities. Apart from finance, it can also be implemented in other areas such as smart contracts, security services, reputation systems, public services, and the Internet of Things (IoT).  Some of the real-world scenarios where we can use blockchain technology are listed below:
\begin{itemize}
	\item \textbf{Peer-to-peer (P2P) global transactions:}  Despite the availability of several international payment processing services like PayPal\cite{PayPal, RealWorldApps}, they often impose significant transaction fees. On the other hand, blockchain provides secure, cheap, and fast transfer of funds across the globe without the need for a trusted third party. Also, various peer-to-peer payment services come with limitations, such as location restrictions and minimum transfer amounts. This is why an increasing number of businesses and individuals now prefer cryptocurrencies for international transfers.
	\item \textbf{Supply chain management and quality assurance:} Blockchain technology provide advantages of traceability and cost-effectiveness in supply chain management. With a blockchain, goods can be tracked, along with their origin, quantity, and other related details. This enhanced level of transparency simplifies various processes in the supply chain ecosystem, such as payments, production process assurance, and ownership transfer. In case of any irregularity detected in the supply chain, a blockchain system can help trace it back to its point of origin. This enables businesses to investigate the issue and take appropriate actions. In the food sector, it is essential to maintain quality and safety by keeping track of vital information such as origin, batch details, and other relevant factors\cite{RealWorldApps}.
	\item \textbf{Accounting:} The use of blockchain technology for recording transactions significantly reduces the risk of human error and safeguards the data against tampering. It's important to note that each time records are passed from one blockchain node to another, they are verified, further ensuring accuracy. This not only provides guaranteed accuracy for your records but also creates a highly traceable record of financial transactions. 
	\item \textbf{Smart contracts:} Lengthy contractual transactions can impede the growth of a business, especially for enterprises that deal with a high volume of communications regularly. Smart contracts enable the automatic validation, signing, and enforcement of agreements through a blockchain framework. This eliminates the need for intermediaries, thereby saving the company both money and time.  A  cryptographer Nick Szabo\cite{NickSzabo}, in 1994, recognized the potential of using the decentralized ledger for smart contracts, also known as self-executing digital contracts using blockchain. With this approach, contracts could be transformed into computer code, stored and duplicated on the system, and overseen by the network of computers that operate the blockchain.
	\item \textbf{Voting:}  Using blockchain for local elections could significantly reduce the risk of electoral fraud, which is a major concern despite the widespread use of electronic voting systems. When NASDAQ (National Association of Securities Dealers Automated Quotations)\cite{NASDAQ} utilized blockchain technology to streamline shareholder voting, it collaborated with its blockchain technology partner and local digital identification solutions that issued identity cards to governments. This initiative termed the ``e-voting'' project, was deemed disruptive, necessary, and practical after it yielded positive results.
	\item \textbf{Stock exchange} Blockchain technology has been considered a potential solution for securities and commodities trading for some time, owing to the dependable yet transparent nature of blockchain systems. This has led stock exchanges to explore its potential as the next significant advancement. As an example of this idea, the ASX (Australian Stock Exchange)\cite{ASX,RealWorldApps} has already made plans to transition to a blockchain-based system for their operations, utilizing technology developed by a blockchain startup Digital Asset Holdings. ASX declared in December 2017 that it chose Digital Asset to develop a new DLT-based (Distributed Ledger Technology)  system to replace its current CHESS (Clearing House Electronic Subregister System) platform. The updated system is expected to bring numerous benefits, such as faster settlement times, enhanced efficiency, and greater security and resilience.
	
	\item \textbf{Energy supply}
	Blockchain technology is now providing sustainable energy solutions that offer precise tracking of usage through ``transactive grids" available for commercial establishments and households in certain regions of the world. For instance, Powerpeers\cite{Powerpeers} in the Netherlands and Exergy\cite{Exergy,RealWorldApps} in Brooklyn are two such examples. Additionally, blockchain can be employed to enhance the monitoring of clean energy.
	
	\item \textbf{IoT devices} Blockchain technology has the potential to offer a secure mesh network for the Internet of Things (IoT)\cite{IOT} to interconnect without the risks associated with central server models. This can create a platform for a communal economy based on machine-to-machine interactions. Through blockchain, data generated from IoT sensors can be monetized, enabling owners of IoT devices to sell such data for digital currency.
	
	\item \textbf{e - Auction} The integration of blockchain technology into e-auctions can improve transparency, security, and efficiency. By adopting a decentralized platform, blockchain eliminates the need for intermediaries and guarantees a secure recording of all transactional data. This enhances trust between buyers and sellers, as all participants have real-time visibility into the entire bidding process, reducing the likelihood of fraudulent activities. Moreover, the implementation of smart contracts in blockchain-based e-auctions can automate the bidding process and ensure that all parties comply with the predetermined terms and conditions. As a result, blockchain technology has the potential to revolutionize the e-auction\cite{eAuction} industry by offering a more transparent, secure, and efficient platform for online auctions. These systems can be enhanced by privacy-preserving primitives as well. In conventional e-auctions, every participant can see the bids submitted by other bidders, potentially resulting in sensitive information being revealed and some bidders avoiding participation. Conversely, in privacy-preserving e-auctions, bids are encrypted, and the identity of bidders remains confidential, guaranteeing the security and privacy of their information. Furthermore, privacy-preserving e-auctions use advanced techniques such as homomorphic encryption and secure multi-party computation to allow the Auctioneer to calculate the winning bid without disclosing individual bids\cite{BernardoMD}.
	
	\item\textbf{NFT (Non-Fungible Token)} NFTs are digital assets stored on a blockchain, usually, the Ethereum blockchain, that are distinct and cannot be divided. Unlike fungible tokens such as Bitcoin or Ether, which can be traded on a one-to-one basis, NFTs are singular and unique. NFTs\cite{NFT, RealWorldApps} are utilized to signify ownership of digital assets like art, music, videos, and other creative content. This enables creators to sell their digital works as one-of-a-kind, valuable items, akin to physical art pieces. Buyers can authenticate ownership and legitimacy of the digital asset using the blockchain.
\end{itemize}
 One can refer to\cite{prewett2018blockchain} for possible future blockchain adaption barriers such as scalability, system integration, lack of standardization, the complexity of blockchain applications, regulatory uncertainty, and risks such as architecture and design risks as well as endpoints, storage, data security and confidentiality, smart contract, compliance, vendor and contractual risks.

\subsubsection{Evolution of Blockchain and Cryptocurrencies in Finance}
Now we briefly look at the historical evolution of digital cash systems from the most primitive forms until today's complicated Bitcoin form. In 1982, David Chaum invented a digital cash and blind signature system\cite{cryptoeprint:2022/026,10.1007/978-1-4757-0602-4_18}. Even though bankrupted in later years, he founded  the electronic cash  company \textit{DigiCash} which allows an untraceable digital payment system based on cryptographic digital signatures. Later in 1997, Adam Black created a  Proof-of-Work algorithm \cite{Back2002HashcashA,cryptoeprint:2022/026} another important concept of blockchain systems with \textit{Hashcash} which was initially created as a denial of service countermeasure  for unwanted emails and internet sources. Soon after in 1998, Nicholas Szabo proposed \textit{Bit gold} which is a simple version of the current Bitcoin  protocol by integrating the Proof-of-Work into a computer network so that it requires solving a cryptographic puzzle and uses Byzantine agreement protocol that is based on the addresses of the majority rather than the computer power\cite{cryptoeprint:2022/026}. Yet again in 1998, Wei Dai proposed \textit{b-money} an untraceable distributed electronic cash which is also used to enforce contracts within a group of pseudonyms\cite{cryptoeprint:2022/026, Bmoney}. Finally, in 2008, Bitcoin a peer-to-peer (P2P) electronic cash system was introduced by the alias name author(s) Satoshi Nakamoto. The proposed system suggests a solution for the double spending problem as well\cite{cryptoeprint:2022/026, nakamoto2008bitcoin,IOTA}. Then in 2009, the first Bitcoin network technology is constructed as a decentralized ledger. This new financial system that comes with Bitcoin 1.0 stores the transactions in an immutable, decentralized, and distributed way in the blockchain nodes. Later in 2013,  Ethereum  which is a blockchain with smart contract functionality was conceived by Vitalik Buterin. In current market capitalization \textit{Ether} is the second only to \textit{Bitcoin}\cite{Ethereum,cryptoeprint:2022/026}. 

\subsubsection{Consensus Mechanism in Blockchain} Consensus is a flexible approach for achieving agreement within a group that considers the perspectives and interests of all members. In contrast to voting, which may only consider the majority opinion, consensus seeks to reach a solution that benefits the group as a whole. There is more than one type of consensus in blockchain such as consensus in the state, consensus in payment, and consensus in rules for the network. Prior to the advent of Bitcoin, numerous attempts at creating peer-to-peer decentralized currency systems had failed due to their inability to solve the most significant challenge in achieving consensus, known as the ``Byzantine Generals Problem". The generals require a consensus mechanism that ensures their army can operate as a unified force in the face of any obstacles or challenges\cite{lamport1982byzantine}. For instance, when sending 7 Ether from your wallet to someone else, how can you be certain that the transaction amount won't be altered by a third party in the network, increasing it to 70 Ether?  Consensus mechanisms deal with such problems. To ensure the transaction amount remains tamper-proof during its transmission, it will be broadcasted to the network and verified by the nodes in the network through a consensus mechanism. Once a sufficient number of nodes have verified the transaction, it will be recorded on the blockchain, where it cannot be altered or deleted without consensus from the majority of nodes on the network. This makes it extremely difficult, if not impossible, for someone to tamper with the transaction amount without detection. Therefore, the receiver can be confident that the amount they receive is the exact amount sent. Below given a list of some consensus mechanisms:
\begin{itemize}
	\item 	\textbf{Proof of Work (PoW):} In this system, to add a block to the blockchain, special nodes called ``miners", must solve complex cryptographic hash puzzles that demand significant computational power and energy. These puzzles are intentionally designed to be challenging on the system. Once a miner solves the puzzle, they submit their block to the network for verification, which is a straightforward process to determine if the block is valid or not. One disadvantage of the Proof-of-Work mechanism is that it is highly inefficient due to the significant power and energy consumption it requires. Individuals or entities that can afford to acquire faster and more powerful ASICs have a higher probability of successful mining compared to those who cannot. This issue can risk the decentralization concept. Bitcoin (BTC), Ethereum Classic (ETC), Litecoin (LTC), Monero (XMR), and Bitcoin Cash (BCH) are examples of blockchains using the PoW consensus protocol.
	\item\textbf{Proof of Stake (PoS)}: In the Proof of Stake consensus mechanism, there is no need for solving complex mathematical puzzles. Instead, the creator of a new block is selected in a deterministic way based on their ``stake" in the network. Stake refers to the number of coins or tokens one possesses. In that sense, the higher the stake, the greater the probability of being chosen as the next block validator. A significant benefit of the Proof of Stake system is its increased energy efficiency, as it eliminates the energy-intensive mining process, making it a potentially greener alternative to Proof-of-Work systems. On the other hand, a disadvantage of PoS is it provides a motivation for large stakeholders to amass more cryptocurrency in order to enhance their chances of being selected as validators and earning rewards which may cause centralization. Cardano (ADA), Casper, and Algorand (ALGO) are examples of cryptocurrencies that uses a PoS type of consensus mechanism.
\end{itemize}
Some other consensus algorithms can be listed as: \begin{itemize}
\item	Practical Byzantine Fault Tolerance (PBFT), Delegated Proof of Stake (DPOS), Proof of Importance (PoI), Proof of Burn (PoB), Proof of Luck (PoL), Delegated Byzantine Fault Tolerant   (dBFT), Proof of Concept (PoC), Proof of Exercise (PoX), Proof of Capacity, Proof of Elapsed Time, Proof of Weight, and many more.
\end{itemize}

\subsubsection{Cryptographic Aspects of Blockchain} Cryptography is the main underlying tool of blockchain \textbf{security} and \textbf{privacy}. The primary cryptographic primitives of blockchains are \textbf{hash functions} and \textbf{digital signature algorithms} both of which provide resistance to unauthorized alterations and modifications on the ledger, attainment of consensus, and public verifiability.  
\subsubsection*{Digital Signature Algorithms}
Digital signature algorithms are based on the public key cryptographic systems. It is used to sign the transactions with the signer's private key and the verification is done by the public key by the other nodes. They provide \textit{authenticity} of the transaction source, \textit{data integrity} i.e. the data is not being tampered with by a malicious attacker, and \textit{non-repudiation}. Non-repudiation is the ability to prove that a certain action or transaction has been performed by a specific party and that this party cannot deny having performed the action or transaction. In other words, it is the assurance that the originator of a message or a transaction cannot deny their involvement in it. 
The most commonly used digital signature algorithms are:
\begin{itemize}
\item{\textbf{RSA (Rivest-Shamir-Adleman Digital Signature Algorithm)}}: The RSA\cite{RSA1} encryption method is a commonly employed public-key cryptographic system utilized for ensuring secure data transmission and digital signatures. In the RSA digital signature algorithm, messages are signed using a private key, and the signature can be verified using a corresponding public key. The underlying cryptographically hard problem of the RSA signature scheme is the Integer Factorization Problem (IFP)\cite{FactoringRSA2019}. Some cryptocurrencies that use the RSA signature scheme can be listed as Namecoin, Hedera, and Arweave. Although it is not always the main digital signature algorithm in some coins like Bitcoin, Litecoin, and Dogecoin,  RSA is utilized in some aspects of those protocols, including key generation and signature verification for multi-signature transactions (Bitcoin), encryption and decryption of private keys (Dogecoin), or generating and verifying SSL certificates for websites hosted on its decentralized domain name system (Namecoin).
\item \textbf{ECDSA (Elliptic Curve Digital Signature Algorithm)}:  This signature scheme is based on the mathematically hard Discrete Logarithm Problem (DLP) \cite{DL1,DL2} on the elliptic curves. ECDSA has an advantage over RSA (Rivest-Shamir-Adleman) in a way that it can provide equivalent security with smaller key sizes. As a result, ECDSA is more efficient in terms of storage and transmission, since smaller keys require fewer computational resources to process.

Some of the popular cryptocurrencies that use ECDSA can be listed as -  Bitcoin (BTC), Ethereum (ETH), Litecoin (LTC), Bitcoin Cash (BCH), Ripple (XRP), Binance Coin (BNB), Cardano (ADA), Polkadot (DOT), Chainlink (LINK), Dogecoin (DOGE).
ECDSA is a widely used digital signature algorithm, as it provides strong security while using smaller key sizes than other traditional digital signature algorithms.
\item \textbf{EdDSA (Edwards Curve Digital Signature Algorithm)}:

While not as widely used as ECDSA, some cryptocurrencies and blockchain networks have adopted EdDSA for their digital signature algorithm. Here are a few examples: Monero (XMR), Zcash(ZEC), Stellar (XLM), Nano (NANO), Grin (GRIN), and Solana (SOL).
EdDSA provides strong security and faster signature generation than the other digital signature algorithms, which makes it an attractive choice for blockchain networks. 
\end{itemize}

 There exist other different types of  digital signature algorithms with several advantages such as providing \textbf{``privacy''}. Privacy can be achieved when some digital signature algorithms are used in combination with relevant  cryptographic primitives. Some of these types of signature algorithms can be listed as \textbf{ring signatures}, \textbf{blind signatures}, \textbf{threshold signatures}, \textbf{multi-signature schemes} and \textbf{zero-knowledge proof commitments}\cite{BernardoMD,cryptoeprint:2022/026}.  Below given are some different types of signature schemes that are used in various blockchains to achieve certain privacy-related attributes.
  
\begin{itemize}
	\item 	\textbf{Multi-Signatures:} A multi-signature scheme\cite{yao2006secure} is a cryptographic protocol that enables a group of individuals to collectively approve a transaction or gain access to a resource. To do so, a specific number of private keys are necessary to authorize the transaction or access request. These private keys are distributed among the parties involved, and all parties must provide their approval by signing off using their private key. Several cryptographic techniques such as multi-party computation, threshold encryption, homomorphic encryption, and secret sharing schemes are used in the construction of multi-signatures. 
	\item \textbf{Threshold-Signatures:} A threshold signature\cite{camenisch1997threshold,boldyreva} is a digital signature that enables a group of signers to collaboratively sign a message, requiring only a minimum number $t$  (known as the threshold) out of a total of $n$ signers to cooperate to generate a valid signature. The threshold, established during setup, sets the minimum number of signers necessary to agree on the signature to make it valid. A threshold signature scheme is more secure than traditional signatures because it requires the cooperation of a minimum number of signers to produce a valid signature. This reduces the risk of a single compromised device or malicious signer being able to sign a message. Threshold signatures can be more efficient than traditional signatures, as they reduce the need for multiple rounds of signing and key exchange. This can be especially useful in resource-constrained environments.  Threshold signatures distribute the secret key among multiple parties, which can help to mitigate the risk of key exposure.  In general, while threshold signatures do not provide anonymity on their own, they can be combined with other techniques to achieve strong anonymity guarantees in various applications. 
	\item 	\textbf{Schnorr Multi-Signatures:} Schnorr signature is a type of multi-signature scheme that was introduced by Claus-Peter Schnorr in 1991 \cite{schnorr1991efficient,maxwell2018simple}. The main advantage of  this multi-signature scheme over the standard multi-signature ones is that it allows key aggregation and batch verification which helps multiple signatures to be verified simultaneously, resulting in faster verification times. Another advantage is that is \textit{provable secure}\cite{rivest1978method} meaning that can be mathematically demonstrated to be secure without relying on any unproven assumptions about the fundamental cryptographic building blocks.
	\item 	\textbf{Ring-Signatures:} In a ring signature scheme\cite{rivest2001leak}, a user can sign a message anonymously on behalf of a group, known as a ring or a set, with the identity of the signer kept concealed and only the members of the ring able to authenticate the signature. Ring signatures are highly beneficial due to their robust anonymity guarantees. Since any member of the ring can produce the signature, it becomes infeasible to ascertain the actual identity of the signer. As a result, ring signatures find application in various use cases, such as cryptocurrencies, voting systems, and whistleblowing, among others. While other types of signatures can offer anonymity, only ring signatures are utilized to provide the anonymity\cite{raikwar2019survey} of the signer in blockchain transactions. Additionally, they can be used to create untraceable payments\cite{saberhagen2013cryptonote,wang2019cryptographic,bender2005ring,chen2006traceable,traceable2}. 
	
	\item 	\textbf{Blind-Signatures:} A blind signature is a digital signature scheme that enables a message to be signed by a signer without the signer seeing its content. The blind signature protocol is used to allow a user to obtain a signature on a message or a transaction in the cryptocurrencies from a signer without disclosing the message's content to the signer. The blind signatures provide anonymity and unlinkability\cite{cryptoeprint:2022/026,heilman2016blindly,chaum1982blind,raikwar2019sok}.
	\item 	\textbf{ZK-SNARKs (Zero-Knowledge Succinct Non-Interactive Argument of Knowledge):}  ZK-SNARK is utilized in which one party can demonstrate to another party that they possess certain information without disclosing that information or any interaction between the parties. The use of ZK-SNARKs is common in several blockchain applications, particularly privacy coins, as they enable transaction verification without exposing the parties and transaction amounts involved\cite{zcash,bitansky2011extractable,ben2014zerocash}. Discrete Logarithm Problem (DLP)\cite{DL1,DL2} is the underlying cryptographically hard problem of ZK-ZNARKs\cite{kearney2021vulnerability,cryptoeprint:2022/026}. ZCash\cite{EF} is an example of this scheme, launched in 2016, which is a cryptocurrency that emphasizes privacy. Although it is built on the same underlying code as Bitcoin, ZCash includes an additional privacy layer that enables users to conduct transactions without disclosing either their identities or the transaction amount. ZCash provides two categories of addresses: \textit{transparent} and \textit{shielded}. While transparent addresses are akin to those of Bitcoin and are observable on the blockchain, shielded addresses, leveraging ZK-SNARKs, are entirely confidential, preventing any transaction information from being visible on the blockchain. As a result, ZCash has become a popular option for those who value privacy and security in their financial transactions, attracting both individuals and corporations.
	\item 	\textbf{Bulletproofs:}  The Bulletproofs\cite{bunz2017bulletproofs} signature scheme is also the type of a zero-knowledge proof system that is non-interactive, meaning that it allows one party to prove the truth of a statement to another party without disclosing any additional information beyond the veracity of the statement. Bulletproofs is a newer advancement in zero-knowledge proof systems that aim to improve efficiency over ZK-SNARKs, particularly in terms of proof generation time and the size of the proofs. Unlike ZK-SNARKs, Bulletproofs do not require a trusted setup and utilize a distinct mathematical approach to generate proofs that are more compact and quicker to validate. Monero (XMR), Grin (GRIN), and Beam (BEAM) can be examples of coins that use a bulletproof scheme.
	
	Apart from all of these digital signature schemes, there exists \textbf{post-quantum digital signatures schemes}, constructed in such a way that they provide security against  possible attacks by quantum computers, which we will discuss later.			
\end{itemize}

\subsubsection*{Hash Functions}

A hash function is a type of mathematical function that accepts an input (referred to as a message, data, or transaction) of any length and produces a fixed-length output known as a hash value or digest. The hash function is intended to be a one-way function, implying that computing the hash value from the input is simple, but producing the original input from the hash value is extremely difficult or practically impossible. Examples of widely used hash functions are \textit{SHA-256 (Secure Hash Algorithm 256-bit), SHA-3 (Secure Hash Algorithm 3), BLAKE2, and RIPEMD (RACE Integrity Primitives Evaluation Message Digest)}. Some popular hash functions that are used in cryptocurrencies are:
\begin{itemize}
\item\textit{SHA-256 (Secure Hash Algorithm 256-bit)}: Bitcoin and many others.
\item \textit{Scrypt}: Litecoin, Tenebix, Fairbix, and many others. 
\item \textit{Ethash}: Ethereum and Ethereum-based cryptocurrencies.
\item \textit{Blake2b}: Siacoin and many others.
\item \textit{Equihash}: Zcash and many others. 
\item \textit{X11}: Dash, Darkcoin and others.
\item \textit{CryptoNight}: Monero  and many others.
\end{itemize}

 Hash functions find application in blockchain by solving cryptographic puzzles through the mining (PoW) process, generating addresses for public and private keys, creating blocks using the Merkle-tree paradigm (MKT), and generating pseudorandom numbers (PNG)\cite{cryptoeprint:2022/026}.

\subsubsection*{Bitcoin Mining (PoW) and Hash Function} Bitcoin uses \textit{SHA256d} hash function. During the mining process, in order to add a new block to the blockchain, the miner must discover the \textbf{nonce} (a 32-bit field random number used once in a block header that is incremented by miners until the hash of the block header meets a certain difficulty target) that satisfies the following puzzle: $SHA256d(X||nonce) \leq n$, where $n$ is the target value of $256$ bits. The first $k$ blocks of the hashed value must consist of all zeros, and $k$ is adjusted after every $2016$ block generated to keep the average time for the generation process around 10 minutes. The primary purpose of PoW, based on Back's concept from 1997, is to enable a decentralized group without pre-established trust to reach a consensus on a consistent transaction history and prevent double-spending attacks\cite{cryptoeprint:2022/026}.

.

\subsubsection*{Blockchain Forking} A fork in a blockchain occurs when the state of the blockchain splits into separate chains due to different segments of the network holding contrasting views on the transaction history. In essence, a fork represents a discrepancy in the interpretation of the blockchain's state. When multiple nodes in a decentralized network discover a suitable nonce almost simultaneously, it is possible for valid blocks to be generated concurrently, resulting in the formation of branches. To address this, each node in the network always strives to select and extend the chain of blocks that exhibits the highest Proof-of-Work (PoW), which is referred to as the longest chain or the chain with the greatest cumulative difficulty. There are two basic types of forks:
\begin{itemize}
\item \textbf{Soft Fork:} A soft fork refers to a situation where changes are made to the blockchain protocol that can be integrated into previous versions of the software without disrupting the existing network. This implies that nodes operating on the older version of the software will still consider new blocks as valid, and the blockchain will remain unified. Usually, soft forks are implemented to incorporate new functionalities or enhance the blockchain's efficiency without causing a complete division in the network. Given that nodes running previous software versions can still function in the network, a hard fork that demands all nodes to upgrade to the latest software version to maintain compatibility with the blockchain is unnecessary. For instance  \textbf{Segregated Witness (SegWit) }is a soft fork that was implemented on the Bitcoin blockchain to enhance network efficiency and transaction capacity. The SegWit update involves separating the signature data from the transaction data, thereby freeing up more space in each block and allowing for a higher volume of transactions to be processed at once. \textbf{BIP65, BIP66, and  P2SH - Pay-to-Script-Hash (P2SH)}  are also other examples of Bitcoin soft forks. In all of these instances, the modifications made to the blockchain protocol were designed to be backward compatible, which means that previous software versions can still participate in the network without experiencing any interruptions or difficulties.
\item \textbf{Hard Fork:}  Hard forks are fundamental changes in blockchains that are not backward compatible, unlike the case of soft forks. When a hard fork is implemented, the blockchain protocol undergoes a significant transformation that is incompatible with previous software versions. This results in older nodes being unable to validate new blocks, leading to the creation of two separate chains. Typically, hard forks arise due to disagreements among network participants regarding the blockchain's operational rules, such as modifications to transaction fees or block size. All nodes must upgrade to the new software version to ensure compatibility with the blockchain, and those who fail to do so are left with the old chain, resulting in the development of a new cryptocurrency. \textbf{Bitcoin Cash (BCH)} and \textbf{Bitcoin Gold (BTG)} are examples of hard fork on Bitcoin blockchain whereas \textbf{Ethereum Classic (ETC)} is a hard fork for Ethereum.
\end{itemize}

\subsubsection{Bitcoin and Economics} The rate of block creation for bitcoin is designed to be constant, with six blocks produced per hour, which translates to one block every 10 minutes. This rate is maintained to prevent \textbf{inflation} in the Bitcoin network. However, as the mining hash rate increases over time, the difficulty of the mining problem also increases to keep up with the pace. When Bitcoin was first introduced in 2009, the initial block reward for successful mining was 50 BTC. This reward is halved every four years after every 210,000 blocks have been mined. Currently, the block reward for successful mining is 6.25 BTC. This halving process will continue until around the year 2110-2140, by which time 21 million BTC will have been issued. 

The limited supply of Bitcoin, similar to gold, implies that its value is expected to rise over time. Hence, Bitcoin is often referred to as \textbf{``digital gold"}. Bitcoin's value is expected to increase until the last Bitcoin is mined, at which point it may plateau. While there may be fluctuations in value, Bitcoin's status as the oldest cryptocurrency has led to a recent stabilization. In  deflation type of models, less money is produced each year, with the expectation that the value of fiat currency or Bitcoin will appreciate over time. Bitcoin as a currency can lead to an increase in the overall money supply, which can result in rising price levels (assuming the velocity of money and quantity of goods and services remain constant). However, since Bitcoin is inherently deflationary due to its limited supply, it may also benefit from causing a decline in the value of other currencies while experiencing appreciation in its own value.

 In the past, gold has been widely recognized as a limited commodity and a popular asset to include in one's portfolio to hedge against the potential impact of expansive monetary policies. Similarly, Bitcoin shares some similarities with gold as a finite commodity, although it does not have any intrinsic value. When the value of gold began to decline due to the price cap it had reached and the partial recovery of the financial markets, the value of Bitcoin increased significantly. This was due to the clever design of Bitcoin as a financial product with a limited quantity.

\subsection{Security Vulnerabilities of Blockchain Systems}
Bitcoin is currently being considered as a stock in the financial market\cite{dasgupta2018survey}. It should be noted that blockchain in fact is a computer software that enables transactions among participants in a peer-to-peer network. Thus, important  questions regarding blockchain security are being discussed. In this section, we provide a perspective on the various kind of security concerns on \textbf{quantum threats} surrounding blockchain. Certainly, blockchains  have many more other security vulnerabilities than quantum-related ones. The reviews  \cite{dasgupta2018survey,xia2019blockchain,luo2019security,cain2020systematic}  are good sources for  a comprehensive analysis of  not necessarily only quantum-related but all sorts of security threats against blockchains, such as double spending attack, $\%51$ attack, selfish mining, Sybil attack, replay attack, manipulation based attacks, eclipse attacks, transaction malleability, timejacking, reputation based attacks, race attacks, DDOS attacks, Finney attack, vector76 attack, collusion attack, malware attacks and many more.  However, our main focus will be the cryptographic vulnerabilities, which is more relevant to this paper's purpose i.e. exploring the effects and risks of quantum computing on finance,  since cryptographic primitives within the blockchain structure is the ones most affected by the rising quantum computing threats.

\subsubsection{Cryptographic Vulnerabilities}
 The three basic cryptographic primitives  that are used in blockchain systems  are digital signature algorithms, hash functions, and random number generators. Below we discuss attacks and threats toward each of those primitives separately:
\begin{itemize}
	\item \textbf{Digital Signature Algorithm Vulnerabilities}
	 Digital signature algorithms which are based on public key cryptography are the main security pillars of the blockchain when conducting transactions and keeping track of past activities. Each blockchain uses its own choice of digital signature algorithms and hash functions. Commonly used digital signature algorithms for blockchains  are ECDSA (Elliptic Curve Digital Signature Algorithm), and EdDSA (Edwards Curve Digital Signatıre Algorithm). Each of these algorithms are based on a mathematical hard problem called the Elliptic Curves Discrete Logarithm Problem (ECDLP) which is known to be an NP-hard problem i.e. it is computationally infeasible to find a solution to this problem in a reasonable amount of time.  In other words, if someone wants to break the cryptographic structure of the ECDSA algorithm this person needs to solve the ECDLP NP-Hard problem within a reasonable amount of time and computational sources. Any vulnerabilities in the underlying cryptographic construction of these ECDSA or EdDSA can cause serious security violations. 
	
	Some cryptographers are skeptical about the secp256k1 curve which is used in the ECDSA of Bitcoin and Ethereum due to its poorly explained curve parameter derivation process, which leaves room for potential manipulation. This uncertainty raises concerns that intentional weaknesses or deficiencies that may lead to security vulnerabilities, may exist within the curve\cite{dasgupta2018survey,Bernstein2014}. Furthermore, the operations of addition and doubling in ECC (Elliptic Curve Cryptography) have distinct timing and power consumption characteristics, which can potentially be exploited through side-channel attacks, fault analysis, timing attacks, and power attacks\cite{Stanford}. 
	
		\item \textbf{Hash Function Vulnerabilities} SHA-256 is a secure hash algorithm that is commonly used in blockchain systems and is vulnerable to length extension attack\cite{dasgupta2018survey}. The attack is based on modifying the hash of a signed message or transaction by appending certain data by the attacker to the message without the need to know the shared secret. A countermeasure to this attack is given in\cite{schneier2003practical} as double use of SHA256. Hash algorithms are vulnerable to birthday attacks which are based on finding hash collisions and this attack managed to break the SHA-1 algorithm\cite{stevens2017first}. Because of these attacks, SHA-1 and MD5 hash algorithms are broken and considered no longer secure for cryptographic purposes. 
		
			\item \textbf{Random Number Generator Vulnerabilities} Random number generators (RNGs) play a crucial role in blockchain technology and are utilized for several functions, including the selection of validators or miners, the verification of transactions, and the creation of new blocks. Miners utilize an RNG to produce a random \textbf{nonce (a number that is used once)} that is then combined with the transaction data and hashed with the block header to authenticate transactions. This procedure is repeated until the hash value meets a predefined difficulty threshold, and the miner who discovers the accurate hash value first is compensated with cryptocurrency. Another common use can be seen in PoS-based blockchains. Proof of Stake (PoS) selects validators to create new blocks depending on the quantity of cryptocurrency they possess and to maintain fairness and avert centralization, an RNG is frequently utilized in the selection process. Random number generators RNGs are  also implemented in ECDSA (Elliptic Curve Digital Signature Algorithm) to create a secret nonce that is applied in the signing process. This nonce is a randomly generated value used in the signing algorithm to guarantee that an identical signature cannot be generated for different messages. Their vulnerabilities come with the cryptographically secure  random number generators that the ECDSA and most other cryptographic protocols rely on. A poorly chosen random number generator can lead to recovering the private key from the given public one. An example of this type of attack can be  the  one performed on Android Bitcoin wallets due to the weakness of the pseudo-random number generator of Java\cite{dasgupta2018survey}. Yet another attack comes with using lattices in\cite{breitner2018biased}. According to this study if the nonce  of ECDSA is biased  and some bits are known, then one can recover the private key by using HNP (Hidden Number Problem) and Lattice Reduction Algorithms. Thus, it’s crucial for the security of ECDSA that the nonce be generated by a cryptographically secure  and unbiased random number generator. 
			The vulnerabilities related to the poor pseudo-random number generators can theoretically be overcome by using \textbf{quantum  random number generators} which are true random number generators \cite{shi2019unbiased} and provides the security needed against the bias-related attacks.
			
\end{itemize}

\subsubsection{Quantum Vulnerabilities} It is widely believed that quantum computers\cite{yeniaras2020faster,GoogleAI,IBM-Google,ImprovedEsra} have the potential to solve difficult computational problems, such as the Integer Factorization Problem (IFP)\cite{FactoringRSA2019} and the Discrete Logarithm Problem (DLP)\cite{DL1,DL2}, which are the security foundations of most cryptographic protocols. Shor's quantum factoring algorithm\cite{ShorsFactoring,ShorsFactoring2} makes the most commonly used asymmetric cryptographic systems, such as RSA\cite{RSA1}, Diffie-Hellman\cite{DiffieHellman1,DiffieHellman2}, and Elliptic Curve Diffie-Hellman\cite{ECDH1,ECDH2,ECDH3,ECDH4,ECDH5} protocols, vulnerable to attacks by sufficiently powerful quantum computers. Grover's algorithm\cite{GroversSearch} improves brute force attacks by significantly reducing the search space for private keys, pre-image, and collision attacks to hash functions. These algorithms are used to secure sensitive data such as governmental and military information, emails as well as financial data such  as blockchain transactions and wallets, thus compromising them would have serious consequences for digital security and privacy.  As we discussed in the previous sections, most of the blockchain systems implement  discrete logarithm problem-based ECDSA, EdDSA, or RSA digital signature algorithms and thus all vulnerable to quantum attacks in the near future. Consequently, research and investment in quantum computing have increased in all sectors in recent years, prompting researchers to develop reliable quantum-resistant cryptographic protocols. In the following sections, we will discuss the quantum-proof cryptographic alternatives that can be used in blockchains instead of the ECDSA, EdDSA, or RSA to diminish the quantum attack risks on blockchain systems. Let's briefly describe those quantum algorithms. \\

\textbf{Shor's Algorithm}: Shor's Algorithm\cite{ShorsFactoring,ShorsFactoring2} is  a quantum integer factorization algorithm that significantly surpasses the classical counterpart. Compared to Shor's quantum factorization algorithm, the classical one has exponentially higher time complexity. Shor's algorithm closely resembles the classical approach, as it encompasses a series of classical steps with the exception of one, which is substituted by a quantum algorithm. Let $N$ be an integer such that $N=p.q$ where $p$ and $q$ are two prime numbers. In the most primitive classical method, it takes  $\sqrt N$ trials to find $p$ or $q$. This method is apperantly cumbersome as $e^{(k/2)\ln2}$ for $N=2^k$. Therefore the following algorithm is used instead of the naive method:\\

\textbf{Step 1.} Pick a random integer $n\leq N$ and using Euclidean algorithm, calculate the greatest common divisor gcd(n,N). If $gcd(n,M)\neq1$ then we are done since $n=p$ or $n=q$.\\

\textbf{Step 2 (Quantum Part).} If $gcd(n,N)=1$ then  then  define a function $h:N\mapsto N$ as $x \mapsto n^x\mod N$. Then find the \textbf{order} or \textbf{period} $T$ of $n$ i.e. the smallest integer $T$ which makes $n^T\equiv 1 \mod N$. This step is known to be exponential in time with a classical algorithm but with Shor's quantum algorithm, it takes only polynomial time. Therefore the application of the quantum algorithm using a quantum computer is only needed at this step and all remaining steps can be executed in polynomial time in classical computers. Here are the quantum parts of this step:\\

 Pick an integer $m$ such that $N^2\leq 2^m\leq2N^2$ and define $V=2^m$. Let $h_v:V_m\mapsto V_m$ with $x\mapsto n^x$ where $V_m=\{0,1,...,V-1\}$.\\
 
\textit{Step 2.1.} 
 Configure the registers $r_1$ and $r_2$ to the initial state with $m$ qubits for each register.
 $$|\phi_0\rangle=|r_1\rangle|r_2\rangle=|00...0\rangle|00...0\rangle$$
 
 \textit{Step 2.2.}  Apply Quantum Fourier Transform (QFT) on the first register: 
  $$|\phi_0\rangle=|0\rangle|0\rangle\mapsto|\phi_1\rangle=(1/\sqrt V)\sum_{x=0}^{V-1}|x\rangle|0\rangle$$
 Thus the first register is in a superposition of all states $|x\rangle$ where $0\leq x\leq V-1.$
 
 \textit{Step 2.3.} Assume that the unitary gate interprets the function $h_v$  as $U_{h_v}|x\rangle|0\rangle=|x\rangle|h_v(x)\rangle$ and apply the unitary gate in step 2.2. to get
 $$U_{h_v}|\phi_1\rangle=|\phi_2\rangle\equiv(1/\sqrt V)\sum_{x=0}^{V-1}|x\rangle|h_v(x)\rangle $$
 which shows that in general the two registers are entangled.
 
 \textit{Step 2.4.} Again applying QFT on $|r_1\rangle$ yields
 $$|\phi_3\rangle=(\mathcal{F}\otimes I)|\phi_2\rangle=(1/\sqrt V)\sum_{x=0}^{V-1}\sum_{y=0}^{V-1}\omega^{-xy}_m|y\rangle|h_v(x)\rangle $$
 $$=(1/\sqrt V)\sum_{y=0}^{V-1}|y\rangle|\psi(y)\rangle= (1/\sqrt V)\sum_{y=0}^{V-1}\lVert|\psi(y)\rangle\rVert\cdot|y\rangle\frac{|\psi(y)\rangle}{\lVert|\psi(y)\rangle\rVert} $$
 where $|\psi(y)\rangle=\sum_{x=0}^{V-1}\omega^{-xy}_m|h_v(x)\rangle $
 
  \textit{Step 2.5.} The measurement of $|r_1\rangle$ yields a $y\in V_m$ with the probability $Prob(y)= \dfrac{{\lVert|\psi(y)\rangle\rVert}^2}{V^2}$
  and simultaneously the state collapses to $|y\rangle\dfrac{|\psi(y)\rangle}{\lVert|\psi(y)\rangle\rVert}$
  
  \textit{Step 2.6.} Get the order $T$ out of this measurement result.\\
  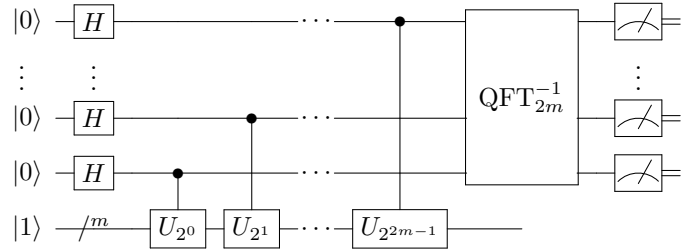
\begin{figure}[!h]
  \centering
   \begin{adjustbox}{width=8.3cm}

  		\Qcircuit @C=.7em @R=.7em {
  			\lstick{\ket{0}}    & \gate{H} & \qw & \qw               & \qw               & \qw & \cdots & & \ctrl{4}               & \multigate{3}{\textrm{QFT}_{2m}^{-1}} & \qw  & \meter & \cw \\
  			\lstick{\vdots\ \ } & \vdots   &     &                   &                   &     &        & &                        &    \pureghost{\textrm{QFT}_{2n}^{-1}} &      & \vdots &     \\
  			\lstick{\ket{0}}    & \gate{H} & \qw & \qw               & \ctrl{2}          & \qw & \cdots & & \qw                    &        \ghost{\textrm{QFT}_{2n}^{-1}} & \qw  & \meter & \cw \\
  			\lstick{\ket{0}}    & \gate{H} & \qw & \ctrl{1}          & \qw               & \qw & \cdots & & \qw                    &        \ghost{\textrm{QFT}_{2n}^{-1}} & \qw  & \meter & \cw \\
  			\lstick{\ket{1}}    & /^m \qw  & \qw & \gate{U_{{2^0}}} & \gate{U_{{2^1}}} & \qw & \cdots & & \gate{U_{{2^{2m-1}}}} & \qw
  		}
  	\end{adjustbox}
  \caption{Period finding quantum circuit design\cite{qcircuit}.}
  \label{fig:shorsfig}
  \end{figure}
 
\textbf{Step 3.} If $T$ is odd, then go back to the first step and repeat until an even period $T$ is found.  When $T$ is even, the following equality holds:
$$(n^{T/2}-1)(n^{T/2}+1)=n^T -1\equiv 0\mod N $$
If $n^{T/2}+1\equiv 0\mod N$ then $gcd(n^{T/2}-1, N)=1$ then go to step 1 and pick another $n$. Assuming that $n^{T/2}-1$ is not a multiple of $N$, if $n^{T/2}+1\not\equiv 0\mod N$ then go to step 4 since in that case $n^{T/2}-1$ includes either $p$ or $q$ as factors thus we proceed to step 4.\\

\textbf{Step 4.} Then either $p$ or $q$ is equal to the $gcd((n^{T/2}-1,N)$  and the factorization is done.\\

In 2001, a team at IBM showcased Shor's algorithm on a quantum computer consisting of 7 qubits\cite{IBMQfactoring}. This demonstration successfully factorized the number 15 into its prime factors, 3 and 5. The qubit requirement for applying Shor's factoring algorithm to cryptographic algorithms is contingent upon the magnitude of the number undergoing factorization. As a general guideline, Shor's algorithm usually necessitates around 2n qubits for efficient factorization of an n-bit number. However, the precise number of qubits can fluctuate depending on the implementation details and optimization methods employed. For instance, when it comes to efficiently factorizing a 2048-bit RSA modulus, a widely utilized component of contemporary cryptography, Shor's algorithm would typically necessitate approximately 4096 qubits. As of now, quantum computers capable of factoring large numbers relevant to cryptography have not yet been constructed. Nevertheless, the field of quantum computing is progressing rapidly, and it remains possible that in a couple of decades that may become possible. Therefore it is critically important to conduct research and development on post-quantum cryptographic algorithms and switch from the classical ones to post-quantum ones until then.\\

\textbf{Grover's Algorithm:} Grover's algorithm\cite{GroversSearch}, developed by Lov Grover in 1996, is a quantum algorithm specifically designed to solve the unsorted search problem or search an unstructured database. It offers a quadratic speedup over classical algorithms. In classical computing, searching an unsorted database typically requires a linear search, which has a time complexity of $O(N)$. In contrast, Grover's algorithm, a quantum algorithm, achieves the fastest possible search time for an unsorted database, operating with a time complexity of $O(\sqrt{N})$. Let's look into the  basic steps of  Grover's search algorithm:\\

\textbf{Step 1.} Select a randomly chosen value that you wish to search for among the qubits.\\

\textbf{Step 2.} Apply the Hadamard gate (H) to induce a superposition state across all the qubits: 
$$|s\rangle =\dfrac{1}{\sqrt N}\sum_{x=0}^{N-1}|x\rangle\ $$

\textbf{Step 3.} Create an oracle, that alters the amplitude of the target object being searched by flipping it.\\

\textbf{Step 4.} Build the diffuser, which performs an inversion operation around the mean 
$$U_s=2|s\rangle \bra s|-I$$

Execute this Grover's iteration (step 3 and step 4) $t(N) $ times.\\
 
\textbf{Step 5.}  Finally, perform a measurement of the resulting quantum state in the computational basis and verify the states by comparing them with the values in the database.\\

  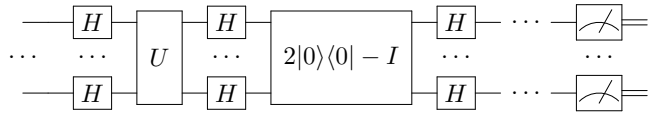
\begin{figure}[!h]
	\centering
 \begin{adjustbox}{width=8.3cm}
\Qcircuit @C=1em @R=.7em {
  \\
	 & \qw & \gate{H} & \multigate{2}{U} & \gate{H} & \multigate{2}{2 \ket{0}\bra{0} - I} & \gate{H} & \qw & \cdots & & \meter & \cw \\
	{\cdots} & & \cdots & & \cdots & & \cdots & & & & \cdots \\
	 & \qw & \gate{H} & \ghost{U} & \gate{H} & \ghost{2 \ket{0}\bra{0} - I} & \gate{H} & \qw & \cdots & & \meter & \cw \\
	 \
}

	\end{adjustbox}
  \caption{A step of Grover's Algorithm quantum circuit design.}
\label{fig:groversfig}

\end{figure}
Grover's algorithm offers significant asymptotic speed improvements for various types of brute-force attacks on symmetric-key cryptography, such as collision attacks and pre-image attacks. It is possible to counteract the potential impact of Grover's algorithm on cryptography, various strategies can be implemented. One method involves increasing the key length utilized in symmetric key encryption algorithms. By doubling the key length, the security level can be sustained against quantum attacks. For instance, if a 128-bit key is deemed secure against classical attacks, a 256-bit key would be advisable to withstand the impact of Grover's algorithm. An alternative approach involves the adoption of post-quantum cryptographic algorithms. These algorithms are explicitly crafted to withstand attacks from both classical and quantum computers. Numerous post-quantum cryptographic schemes are presently under development and standardization, aiming to guarantee the enduring security of encrypted data. Another method is using Quantum Key Distribution (QKD)\cite{QKD1,QKD2,QKD3} which  is a cryptographic method that utilizes quantum mechanics principles to facilitate the secure distribution of encryption keys. QKD enables the establishment of secure keys while also detecting any potential eavesdropping attempts.

\subsection{Post-Quantum Cryptography and Quantum-Safe Blockchain} Post-quantum cryptography (PQC) involves the research and development of cryptographic algorithms that can withstand attacks from quantum computers. As the development of quantum computers progresses, it poses a potential threat to many of the cryptographic algorithms currently in use. Therefore, it is becoming increasingly critical to focus on developing PQC algorithms that can provide long-term security. The National Institute of Standards and Technology (NIST) started a Post Quantum Cryptography Standardization Process in 2016 to evaluate and standardize post-quantum cryptographic algorithms. It is being held as several rounds of competition to which competitors  submit their post-quantum cryptographic  algorithms to be compared and evaluated  from several perspectives such as security vulnerabilities and efficiency. In July 2022, NIST\cite{NISTStatus} announced the selection of 17 post-quantum cryptographic algorithms from the original pool of 69 submissions to advance to the third round of the PQC standardization process. The third round is anticipated to extend throughout 2023 and beyond, with the final selection of standardized algorithms being announced in the future. Those algorithms to be standardized involve Public-key Encryption and Key-establishment Algorithms as well as Digital  Signature Algorithms.

The post-quantum cryptographic algorithms that are submitted to NIST can be divided into five categories, depending on the underlying mathematical hard problem of these algorithms i.e. the mathematical hard problems that makes them quantum-secure for certain sizes of inputs. Below these five main quantum-resistant approaches are listed with examples:
\begin{itemize}
\item \textbf{Lattice-Based Algorithms:}  Security of lattice-based cryptography is based on the hardness of solving certain hard problems such as the Shortest Vector Problem (SVP), Closest Vector Problem (CVP), The Shortest Independent Vector Problem (SIVP), Learning with errors (LWE), Ring Learning with Errors (R-LWE), and Module Learning with Errors (M-LWE). Those hard problems are considered to be quantum-secure in a high-dimensional lattice.  Some lattice-based NIST candidates are  Saber, CRYSTALS-Kyber, CRYSTALS-Dilithium, New Hope, Frodo KEM, NTRU and NTRU Prime\cite{NISTStatus,BernsteinRound1,BernsteinRound2}. Another lattice-based algorithm apart from the NIST candidates is BLISS signature. BLISS (Bimodal Lattice Signature Scheme) is a digital signature algorithm that offers resistance against attacks from both classical and quantum computers. It employs a post-quantum digital signature scheme based on the Short Integer Solution (SIS) and Learning With Errors (LWE) problems in lattices. These mathematical problems are considered hard to solve by both classical and quantum computers, making BLISS a potential candidate for post-quantum digital signature applications.

\item \textbf{Code-Based Algorithms}: Code-based post-quantum cryptographic algorithms rely on the computational infeasibility of decoding specific linear error-correcting codes used to safeguard data from transmission errors, forming the basis of their security. Some code-based NIST PQC candidates are BIKE, Classic McEllice, and HQC.

\item \textbf{Hash-Based Algorithms:} Hash-based post-quantum cryptographic (PQC) algorithms provide security based on the challenge of finding a collision, which is the task of finding two distinct inputs that result in the same hash output, even when employing quantum computers.  Hash-based signature schemes are divided into two groups\cite{buchmann2009hash}. Hash-based One-Time signature schemes (such as Lamport-Diffie\cite{lamport1979constructing} and Winternitz \cite{merkle1989certified}) and those based on the Merkle Signature Scheme (MSS \cite{merkle1989certified}). Numerous adaptations of MSS are discussed in \cite{martos2021white}, including XMSS, XMSS-T, XNYSS, PICNIC, and SPHINCS$^+$\cite{cryptoeprint:2022/026}.

\item \textbf{Super Singular Isogeny-Based Algorithms:} The security of supersingular isogeny-based post-quantum cryptographic (PQC) algorithms stems from the complexity of the isogeny problem, which pertains to discovering the isogeny linking two elliptic curves. These algorithms leverage the assumption that it is difficult, even for quantum computers, to compute the isogeny between two curves, hence providing a reliable foundation for their security. An example NIST PQC candidate is SIKE. However, a preprint was posted on August 5, 2022, by Castryck and Decru stating that there exists an efficient classical key recovery algorithm against Supersingular Isogeny Key Encapsulation (SIKE) and Supersingular Isogeny Diffie-Hellman (SIDH). The preprint included code that demonstrated the practicality of the algorithm. As a result,  the SIKE team acknowledges that these algorithms are now considered insecure and should not be used for post-quantum cryptography\cite{SIKE}.

\item \textbf{Multivariate-Based Algorithms:} Multivariate-based post-quantum cryptography algorithms rely on the premise that it is extremely difficult to solve systems of multivariate polynomial equations. This difficulty is assumed to remain even for quantum computers, making multivariate-based PQC algorithms a potentially suitable option for post-quantum cryptography. Some NIST candidates of this type are GeMSS and Rainbow.
\end{itemize}

\begin{table}[htp]
\centering	
	\label{NISTtable}
\resizebox{0.90\textwidth}{!}{%
	\begin{tabular}{| c | c | c | c | c |}
		\hline
			\textbf{Type} & \multicolumn{2}{ c |}{\textbf{Round 3 Main Candidates}}& \multicolumn{2}{ c |}{\textbf{ Round 3 Alternative Candidates}}  \\ 
			\cline{2-5}
			& \textbf{Algorithm} & \textbf{Category} & \textbf{Algorithm} & \textbf{Category} \\
			\hline
			\multirow{5}{*}{Key Encapsulation Mechanisms}	   &NTRU  & Lattice-based & NTRU Prime  & Lattice-based \\ 
			                                                 &Saber &  Lattice-based &FrodoKEM   &Lattice-based \\ 
			                                                 & Crystals Kyber  & Lattice-based &BIKE  & Code-based\\ 
			                                                 &Classical McEliece & Code-based & HQC & Code-based\\ 
		                                                     &  &   &SIKE  &Isogeny-based \\ \hline
		                                                      
\multirow{3}{*}{Digital Signature Algorithms}	& Crystals Dilithium &Lattice-based  &GeMMS  &Multivariate-based \\ 
			                                      &  Falcon & Lattice-based  & Picnic  & Hash-based \\ 
			                                             & Rainbow &  Multivariate-based &  Sphincs+ & Hash-based\\ \hline
    \end{tabular} %
    }
	\caption{NIST PQC Round 3 Competition Candidates and Alternatives}
 \end{table}
\FloatBarrier

Following thorough deliberation in the third round of the NIST PQC Standardization Process, NIST has recognized four algorithms as potential candidates for standardization. NIST will propose two main algorithms, namely CRYSTALS-KYBER for key-establishment purposes and CRYSTALS-Dilithium for digital signatures, to be implemented in the majority of use cases. Additionally, the signature schemes FALCON and SPHINCS+ will also undergo standardization. Furthermore, The subsequent Key Encapsulation Mechanism (KEM) algorithms have been selected to progress to the fourth round:\\

\noindent- BIKE (Public-Key Encryption/KEM)\\
- Classic McEliece\\
- HQC\\
- SIKE\\

For a detailed and comparative review of all those post-quantum digital signature algorithms from efficiency, security, and size perspectives one can refer to \cite{FernndezCarams2020TowardsPB} as it is a practical and very useful reference.
\subsubsection*{Quantum-Resistant Blockchain}

A number of blockchain initiatives are currently focusing on creating solutions that are resistant to quantum computing attacks, in order to ensure the security of their networks over the long term. To construct a cryptographically quantum-resistant blockchain, one needs  a post-quantum digital signature algorithm and a hash function. A post-quantum digital signature algorithm can be chosen from the above-listed ones. Note that, an important finding, presented by Bennett, Bernstein, Brassard, and Vazirani in\cite{bennett1997strengths}, asserts that quantum computing cannot offer an exponential advantage when it comes to searching problems. This means that all symmetric encryption and hash algorithms remain secure against quantum attacks, as it is computationally infeasible to carry out brute force searches for keys and collisions under such circumstances\cite{cryptoeprint:2022/026}.

\begin{table}[htp]
	
	\setlength{\tabcolsep}{3pt}
	\centering	
	\caption{ Some Blockchains with quantum-resistant digital signature algorithms}
	\label{tab:1}
	\resizebox{0.80\textwidth}{!}{%
		\begin{tabular}{|l|l|}
			\hline
			Blockchain/Coin/Cryptocurrency &  Quantum-Resistant Digital Signature Algorithm \\ 
			\hline
			QRL   & XMSS \\ 
	     	IOTA & WOTS (Winternitz one-time signature scheme) \\ 
	    	Mochimo & WOTS (Winternitz one-time signature scheme)\\
			Nexus& FALCON \\
			Hcash& BLISS \\
			Quantum Resistant Coin (QRC)& XMSS \\
			QAN &  XMSSMT (hybrid) \\
			enQlave& XMSS\\
			
			\hline
		\end{tabular}%
	}
	
\end{table}
\FloatBarrier
   \noindent Other than the ones represented in Table~\ref{tab:1}, there exist some research papers that propose different constructions of quantum-resistant blockchains: 
    \begin{itemize}
    	\item In 2018 a quantum-resistant blockchain based on a generalization of BLISS signature is proposed as a  privacy-preserving, one-time linkable ring signature(L2RS), providing  ring confidential transactions on blockchain \cite{OneTimeLinkable}. They name this new cryptocurrency privacy-preserving protocol - Lattice RingCT v1.0.
    	
    	\item There is another study which uses quantum-resistant SPHINCS$^{+}$ signature algoritm in an efficient way in \cite{Sphincs}. 
    	
    	\item The main challenge faced by lattice cryptosystems is that the public keys and signatures they use are typically very large, which limits the number of transactions that can be accommodated in each block of a blockchain. This can significantly impact the blockchain's running speed and performance. To address this issue, Zhang \textit{et al.} in \cite{Zhang2021ABS} proposed a solution where they only store the hash values of public keys and signatures on the blockchain, while storing the complete content of these values on an IPFS which refers to the interplanetary file system. By doing so, the number of bytes required for each transaction is significantly reduced. They have developed a Bitcoin exchange scheme to evaluate the performance of this quantum-resistant blockchain system.
     
    	\item Yet another post-quantum blockchain proposal is introduced in \cite{brindha2022quantum} for which FALCON is used since it is considered to be time efficient and has a smaller key size compared to the other signature algorithms of NIST PQC competition.
    	
    	\item In 2018, Gao \textit{et al.} put forth a cryptocurrency based on a post-quantum blockchain that could withstand quantum computing attacks\cite{Gao}. Nonetheless, their proposition mandated a specialized blockchain and was not harmonious with currently used blockchains\cite{Goodbye}.
    	
    	\item Torres \textit{et al.} presented a post-quantum linkable ring signature (PQLRS) scheme in 2020, which allowed auditors to sign confidential transactions in a distributed fashion\cite{Torres2020PostQuantumLR}. Although their proposal was innovative and appealing, the PQ-LRS was restricted by the size of the signature and the communication overhead it produced\cite{Goodbye,Torres2020PostQuantumLR}.
    	
    	\item In 2020, Shahid \textit{et al.} proposed a distributed ledger that was suitable for financial transactions carried out by lightweight devices (IoT devices). Their approach employed hash-based one-time signatures (OTS) that were secure against quantum attacks\cite{Shahid2020PostquantumDL}.
    	
    	\item  Li \textit{et al.} proposed a signature scheme based on lattices for post-quantum blockchains in\cite{Li2019ANL}. Their approach involved generating keys using Bonsai Trees\cite{cryptoeprint:2022/026}. 
    	
    	\item Chen \textit{et al.} proposed a post-quantum blockchain for smart cities, a lightweight quantum-resistant system that uses an identity-based multivariate-quadratic signature scheme called ID-Rainbow for transactions\cite{Chen2020OnTC,Chen2019IdentityBasedSS,cryptoeprint:2022/026}. 
    	\item Furthermore Proof-of-Work (PoW) scheme - based on lattices- that utilizes the difficulty of Hermite-SVP, a variation of SVP, has been proposed in\cite{LatticePoW}. 
    	
    \end{itemize}

       \noindent It is worth noting that \cite{Goodbye} introduces a methodology for transitioning from non-quantum-resistant  blockchains to quantum-resistant ones by offering a hard fork based on a Proof-of-Burn (PoB) consensus. Thus, instead of implementing a totally new blockchain, it suggests modifying an existing one to convert it to a quantum-resistant one.

\subsubsection*{Quantum-Secure Blockchain}

In the past few years, there has been a growing focus on protecting blockchain from potential attacks by quantum computers. To achieve this, two main approaches have emerged in the field of study, the first one is quantum-resistant blockchain and the second one is quantum-secure or quantum-safe blockchain \cite{Dragan}. The former utilizes a digital signature algorithm that can withstand quantum computing, but is still primarily theoretical and lacks practical application. Remember that hash functions are already quantum-resistant with increased bit sizes. However, implementing quantum-resistant signature algorithms and strong hash functions is unfortunately not enough to avoid all quantum threats. For each layer of the blockchain, there exist various security risks. The concept of quantum-secure blockchain refers to  securing quantum threats in each layer of the blockchain application. The network layer is responsible for providing the interaction and communication of the blockchain nodes with each other. Their primary role is to ensure the reliability of the network, which is why, in the future, this layer will require the implementation of a quantum network. Nodes play a critical role in the hardware layer as well, since they are the physical devices that connect to the network and contribute to the blockchain consensus. To enhance infrastructure security, it is common practice to limit or prevent node access. Therefore, infrastructure enhancements are necessary to fully establish a quantum-secure blockchain.

\subsection{Quantum Blockchain}  Quantum blockchain can be defined as a distributed, encrypted database that utilizes principles of quantum computation and quantum information theory to ensure decentralization\cite{li2019quantum}. It refers to a specific type of blockchain technology that aims to safeguard against possible risks presented by quantum computers. As quantum computers have the ability to solve intricate mathematical problems at an accelerated pace compared to conventional computers, the security of blockchain networks may be compromised. To counteract this, quantum blockchain incorporates sophisticated cryptographic methods, including quantum key distribution, quantum random number generation, quantum network channels, and quantum information theory to safeguard the network and prevent malicious attacks. Moreover, quantum blockchain may also employ quantum-resistant algorithms and protocols to bolster its security measures. Rajan et al. introduced a quantum blockchain scheme using entanglement in time\cite{rajanquantum,li2019quantum}.  Entanglement in time is when microscopic particles, such as photons, that have never existed together can still become entangled. For more information on various quantum blockchain proposals or to learn more about quantum bitcoin mining one can refer to \cite{bettyO,benkoczi2022quantum,YangQuantumBlock,Edwards,wangquantum,MuReQua}

Similar to the traditional blockchain, quantum blockchain also possesses certain traits, including decentralization. However, the key features of quantum blockchain are its security and efficiency. Maintaining the security of the quantum blockchain is of utmost importance. Quantum key distribution (QKD) and quantum secure direct communication (QSDC) are two methods that can be employed to ensure secure communication between nodes. These methods rely on quantum physics principles to authenticate the network and prevent unauthorized access \cite{Dragan}. To solve weaknesses in digital signature algorithms, the quantum blockchain can implement the quantum digital signature mechanism\cite{Dragan}, which imbues the blockchain with quantum security properties. This method ensures that quantum computers are unable to compromise the security of the quantum blockchain \cite{Dragan}. 

Another characteristic of blockchain technology utilizing quantum computing is its high speed of transaction processing. The adoption of Analog Hamiltonian Optimizers has the potential to decrease transaction time and could greatly impact the wider adoption of Bitcoin and other blockchain applications. Additionally, integrating the Grover algorithm into the broader blockchain framework has the potential to improve the efficiency of the mining process. However, those who possess universal quantum computers would have an unfair advantage in acquiring mining rewards until they become more widely accessible. By the time quantum technology becomes widely available, it may be so extensively used that those without a quantum computer will be unable to gain control of the network, while classical hardware lags behind. In general, the advantages of quantum blockchain over traditional blockchain lie mainly in its security, efficiency, and performance\cite{Dragan}.

\begin{table}[htp]
\centering
\caption{A categorization table of the works that are surveyed}
\label{table:8}
\begin{tabular}{ |p{7cm}||p{7cm}|  }
 \hline
 Topic  & Related work surveyed\\
 \hline \hline

 Blockchain basics and real world applications: cryptocurrencies, e-voting, smart contracts, IoT, NFT, Consensus, PoW, PoS & \cite{PayPal}, \cite{PayPal}, \cite{NickSzabo},
\cite{NASDAQ},
\cite{ASX},
\cite{Powerpeers},
\cite{Exergy,RealWorldApps},
\cite{IOT},
\cite{eAuction},
\cite{BernardoMD},
\cite{NFT},
\cite{prewett2018blockchain}, \cite{cryptoeprint:2022/026,10.1007/978-1-4757-0602-4_18},
\cite{Back2002HashcashA},
\cite{Bmoney},
\cite{nakamoto2008bitcoin,IOTA},
\cite{Ethereum},
\cite{lamport1982byzantine}\\

\hline

 Cryptographic Aspects of Blockchain: Digital Signature Algorithms, RSA, ECDSA, EdDSA& \cite{RSA1},
\cite{FactoringRSA2019},
 \cite{DL1,DL2}\\
 \hline

 Privacy Preserving Blockchains: Multi-Signatures, Threshold Signatures, Schnorr Multi-Signature, Ring Signatures, Blind Signatures, ZK Snarks, Bulletproofs& \cite{yao2006secure},
\cite{camenisch1997threshold,boldyreva},
\cite{schnorr1991efficient,maxwell2018simple},
\cite{rivest1978method},
\cite{rivest2001leak},
\cite{raikwar2019survey},
\cite{saberhagen2013cryptonote,wang2019cryptographic,bender2005ring,chen2006traceable,traceable2},
\cite{cryptoeprint:2022/026,heilman2016blindly,chaum1982blind,raikwar2019sok},
\cite{zcash,bitansky2011extractable,ben2014zerocash},
\cite{DL1,DL2},
\cite{kearney2021vulnerability},
\cite{EF},
\cite{bunz2017bulletproofs}\\
 
 \hline
  Security Vulnaribilities of Blockchain: Digital Signatures Based, Hash Based, Random Number Generator Based & \cite{dasgupta2018survey,xia2019blockchain,luo2019security,cain2020systematic},
\cite{Bernstein2014},
\cite{Stanford},
\cite{schneier2003practical},
\cite{stevens2017first},

\cite{breitner2018biased},
\cite{shi2019unbiased}\\
 
 \hline
 
  Quantum computing based security vulnerabilities& \cite{yeniaras2020faster,GoogleAI,IBM-Google,ImprovedEsra}
\cite{FactoringRSA2019},
\cite{DL1,DL2},
\cite{ShorsFactoring,ShorsFactoring2},
\cite{RSA1},
\cite{DiffieHellman1,DiffieHellman2},
\cite{ECDH1,ECDH2,ECDH3,ECDH4,ECDH5},
\cite{IBMQfactoring},
\cite{GroversSearch}\\
 
 \hline

 Post-quantum cryptography, Quantum-resistant blockchain, Quantum Bitcoin Mining, Quantum Key Distribution, and Quantum-secure blockchain& \cite{NISTStatus}, 
\cite{BernsteinRound1,BernsteinRound2},
\cite{buchmann2009hash},
\cite{QKD1},
\cite{lamport1979constructing},
\cite{merkle1989certified},
\cite{martos2021white},
\cite{cryptoeprint:2022/026},
\cite{SIKE},
\cite{FernndezCarams2020TowardsPB},
\cite{bennett1997strengths},
\cite{OneTimeLinkable},
\cite{Sphincs},
\cite{Zhang2021ABS},
\cite{brindha2022quantum},
\cite{Gao},
\cite{Goodbye},
\cite{Torres2020PostQuantumLR},
\cite{Shahid2020PostquantumDL},
\cite{Li2019ANL},
\cite{Chen2020OnTC,Chen2019IdentityBasedSS},
\cite{LatticePoW},
\cite{Dragan},
\cite{li2019quantum},
\cite{rajanquantum},
\cite{bettyO,benkoczi2022quantum,YangQuantumBlock,Edwards,wangquantum,MuReQua}\\
 
 \hline
\end{tabular}
\end{table}

\subsection{Advantages of Implementing Quantum Blockchain} 

Implementing blockchain systems in quantum computers and quantum programming offers several potential advantages:\\

\textbf{Enhanced Computational Power:} Quantum computers have the potential to outperform classical computers in specific computational tasks. By utilizing quantum computing, blockchain systems can execute complex operations more efficiently, leading to faster transaction validation, improved consensus mechanisms, and enhanced scalability.\\

\textbf{Quantum Cryptography:} Quantum blockchain systems can leverage the principles of quantum mechanics to enhance cryptographic protocols. Quantum cryptography, including techniques like quantum key distribution (QKD), enables the secure exchange of encryption keys. This strengthens the confidentiality and integrity of blockchain transactions, making them more resistant to attacks.\\

\textbf{Improved Consensus Mechanisms:} Quantum computing enables the development of novel consensus mechanisms tailored for quantum blockchain systems. These mechanisms leverage quantum properties such as entanglement and superposition to achieve faster and more secure consensus. Quantum-resistant consensus algorithms can also be implemented to ensure long-term security against potential quantum attacks.\\

\textbf{Quantum Data Analysis: }Quantum computing has the potential to revolutionize data analysis and pattern recognition. Quantum blockchain systems can leverage quantum algorithms to process and analyze large datasets more effectively, providing valuable insights and enhancing decision-making processes. This is particularly advantageous in blockchain applications involving complex data sets like supply chain management or healthcare records.\\

\textbf{Quantum-Specific Applications: }Quantum blockchain systems enable the creation of quantum-specific applications and services. These applications harness the unique capabilities of quantum computers, such as quantum machine learning, quantum simulations, or quantum optimization. By combining the benefits of blockchain technology and quantum computing, innovative solutions can be developed for domains such as finance, healthcare, and logistics.\\

\textbf{Future-Proofing: }Implementing blockchain systems in quantum computers and quantum programming allows organizations to prepare for the future era of quantum computing. As quantum computers become more powerful and prevalent, classical cryptographic algorithms used in traditional blockchains may become vulnerable. Quantum blockchain systems provide long-term security, ensuring the protection of sensitive data and transactions in the presence of quantum adversaries.\\

It's worth noting that the practical realization of these advantages relies on advancements in quantum computing hardware, quantum programming languages, and the development of robust quantum algorithms specifically designed for blockchain applications.

\subsection{Obstacles of Implementing Quantum Blockchain} Implementing a blockchain using quantum programming faces numerous challenges due to the early stage of quantum computing and quantum programming. Here are some significant hurdles:\\

\textbf{Limited Availability of Quantum Computing Resources:}  Quantum computers with an adequate number of qubits and low error rates are still under development. Building and maintaining such quantum computers is complex and expensive, resulting in a scarcity of quantum computing resources. This scarcity makes it difficult to scale quantum blockchain implementations effectively.\\

\textbf{Paradigm Shift in Programming: }Quantum programming requires a distinct mindset and approach compared to classical programming. Quantum programming languages like\textbf{ $Q\#$}, \textit{Qiskit}, or \textit{Cirq} have unique syntax and concepts, such as superposition, entanglement, and quantum gates. Developers must acquire expertise in these quantum programming languages and grasp the fundamental principles of quantum mechanics to design and implement quantum blockchain systems effectively.\\

\textbf{Designing Quantum Algorithms: }Creating quantum algorithms tailored specifically for blockchain applications poses a significant challenge. Existing classical blockchain algorithms and protocols cannot be directly transferred to quantum computers due to fundamental differences in computational capabilities and the inherent limitations of quantum systems. Researchers need to explore new algorithms that leverage quantum properties to achieve quantum-enhanced blockchain functionalities, including quantum consensus algorithms or quantum-resistant cryptography.\\

\textbf{Quantum Error Correction: }Quantum computers are susceptible to errors caused by decoherence and noise. Employing robust quantum error correction techniques is crucial to mitigate these errors and maintain reliable and stable computations. However, implementing such error correction mechanisms in quantum blockchain systems is challenging, requiring substantial computational resources and specialized knowledge.\\

\textbf{Integration with Classical Systems: }Quantum blockchains will likely need to interact with existing classical systems and infrastructure. Bridging the gap between classical and quantum systems introduces additional complexities. Ensuring compatibility, data interchange, and secure communication between quantum blockchains and classical components present significant technical challenges.\\

\textbf{Lack of Standardization: }Quantum programming is a rapidly evolving field lacking standardization in programming languages, libraries, and tools. This absence of standardization hampers the development of a cohesive and well-supported quantum blockchain ecosystem. Additionally, the absence of established best practices and guidelines makes it more challenging for developers to create reliable and interoperable quantum blockchain solutions.\\

Despite these obstacles, ongoing research and development efforts are focused on addressing these challenges and advancing quantum programming for blockchain applications. As the field progresses, improvements in quantum computing resources, programming frameworks, and algorithms can be expected, facilitating the implementation and integration of quantum blockchains in the financial system.

\section{Conclusion}
The research in portfolio optimization and application of quantum computing for the same is picking pace, with many banks and quantum computing companies taking a great interest in improving the research in the field as it promises great economic benefits. Current research focuses on finding algorithms better suited to the NISQ era, which can efficiently handle local minima problems and scaling problems, being able to handle combinatorial, convex, or non-convex natures of problems that are induced mainly due to the type of constraints faced by the investors. Algorithms that can handle many constraints are better suited to face real-life problems. 

While the application of quantum computing to optimization problems or the field of machine learning could already bring an advantage in the NISQ era, this will probably not be the case for Monte Carlo methods. The algorithms used here generally require both many qubits and deep circuits.  However, as we have also pointed out, there are numerous ideas and attempts to restrict these requirements fairly. The use of quantum algorithms for derivatives valuation will be particularly important for large institutions and those active in investment banking, as they act as providers of more complex derivatives. The possibility of making risk calculations faster with quantum algorithms has a potentially much broader reach as it affects all banks. In this respect, it can be assumed that more intensive research activities will begin in this area in the next few years. We also review a few of the most recent works on quantum applications for fraud detection. As we can see  quantum computing has the potential to revolutionize the field of fraud detection by enabling faster and more efficient analysis of large datasets. While quantum computing is still in its infancy and some challenges need to be addressed, such as the development of reliable quantum hardware and algorithms, the future looks promising for the use of quantum computing in fraud detection. As more research and development are conducted in this field, quantum computing will likely play an increasingly important role in preventing and detecting fraud in a variety of industries.

Moreover, we discuss one of the most important fintech concepts blockchain and cryptocurrencies from the quantum computing point of view. There have been security issues based on the development of quantum computing that comes with Shor's factoring algorithm as well as  Grover's search algorithms. Although a sufficiently efficient quantum computer has not been constructed yet to place those attacks on the well-known cryptographic protocols, it is a very close threat for the near future. With the store now and decrypt later strategy even today's sensitive information is in danger for blockchain-based financial applications in finance. This paper not only points out the quantum computing-related attacks on blockchain but also  mentions several non-quantum attacks and provides a wide range of  blockchain security-related research papers' review for researchers, fintech developers, and entrepreneurs. We make a comprehensive overview of countermeasures  to take against quantum-related attacks by introducing post-quantum cryptography and quantum-resistant blockchain concepts and the underlying post-quantum digital signature algorithms technology that come into consideration after NIST started the Post-Quantum Cryptography(PQC) Standardization Process in 2016. Yet we make a very detailed analysis of both quantum-resistant and privacy-preserving blockchain applications that implement the zero-knowledge proof systems i.e. ZKSNARks, threshold signatures, multi signatures, ring signatures, bulletproofs, and blind signatures. After clarifying the difference between that quantum-resistant blockchain and quantum-secure blockchain, we finally go over the most recent work on quantum blockchain and quantum mining. As we know quantum computing has the potential power of providing efficiency in bitcoin mining as well as can give a chance to construct a true random number generator that boosts the randomness factor of current pseudo-random number generators that are embedded in classical blockchain applications. We also mentioned the security concerns of hash functions which are another building block of blockchain systems. Finally, after discussing the cryptographic aspects we give a brief overview of the pros and cons of transitioning to a quantum blockchain system and the difficulties of this near-future transformation. 

In conclusion, we have covered the most prominent applications of finance in the quantum computing perspective by reviewing the most recent studies in the field and providing a good state-of-the-art survey for inquisitive researchers who works in near-future quantum technology-related areas.  This can also help fintech application developers, banks, and all other financial institutions to launch new project ideas for development in the field. 

\FloatBarrier

\bibliographystyle{plain}
\bibliography{References}

\end{document}